%% file: outflows.tex
\documentclass[twocolumn]{aastex62}
\usepackage{savesym}
\savesymbol{tablenum}
\usepackage{booktabs}
\usepackage{siunitx}
\restoresymbol{SIX}{tablenum}
\usepackage{amssymb}
\usepackage{amsmath}
\graphicspath{{./}{figures/}}
\received{January 1, 2018}
\revised{January 7, 2018}
\accepted{\today}
\submitjournal{ApJ}
\shorttitle{COMs in Outflows}
\shortauthors{Holdship et al.}
\begin{document}

\title{Observations of CH$_3$OH and CH$_3$CHO in a Sample of Protostellar Outflow Sources}

\correspondingauthor{Jonathan Holdship}
\email{jrh@star.ucl.ac.uk}

\author[0000-0003-4025-1552]{Jonathan Holdship}
\affiliation{Department of Physics and Astronomy, University College London, Gower Street, London WC1E 6BT}

\author[0000-0001-8504-8844]{Serena Viti}
\affiliation{Department of Physics and Astronomy, University College London, Gower Street, London WC1E 6BT}

\author{Claudio Codella}
\affiliation{INAF, Osservatorio Astrofisico di Arcetri, Largo E. Fermi 5, 50125 Firenze, Italy}
\affiliation{Univ. Grenoble Alpes, Institut de Planétologie et d'Astrophysique
de Grenoble (IPAG), 38401 Grenoble, France}

\author{Jonathan Rawlings}
\affiliation{Department of Physics and Astronomy, University College London, Gower Street, London WC1E 6BT}

\author{Izaskun Jimenez-Serra}
\affiliation{Centro de Astrobiología (CSIC, INTA), Ctra. de Ajalvir, km. 4, Torrejón de Ardoz, E-28850 Madrid, Spain}
\nocollaboration

\author{Yenabeb Ayalew}
\affiliation{Hammersmith Academy, London, UK}

\author{Justin Curtis}
\affiliation{Hammersmith Academy, 25 Cathnor Road, London, W12 9JD}

\author{Annur Habib}
\affiliation{Hammersmith Academy, 25 Cathnor Road, London, W12 9JD}

\author{Jamel Lawrence}
\affiliation{Hammersmith Academy, 25 Cathnor Road, London, W12 9JD}

\author{Sumaya Warsame}
\affiliation{Hammersmith Academy, 25 Cathnor Road, London, W12 9JD}

\author{Sarah Horn}
\affiliation{Hammersmith Academy, 25 Cathnor Road, London, W12 9JD}

\begin{abstract}
Iram 30-m Observations towards eight protostellar outflow sources were taken in the 96-\SI{176}{\giga\hertz} range. Transitions of CH$_3$OH and CH$_3$CHO were detected in seven of them. The integrated emission of the transitions of each species that fell into the observed frequency range were measured and fit using RADEX and LTE models. Column densities and gas properties inferred from this fitting are presented. The ratio of the A and E-type isomers of CH$_3$OH indicate that the methanol observed in these outflows was formed on the grain surface. Both species demonstrate a reduction of terminal velocity in their line profiles in faster outflows, indicating destruction in the post-shock gas phase. This destruction, and a near constant ratio of the CH$_3$OH and CH$_3$CHO column densities imply it is most likely that CH$_3$CHO also forms on the grain surface.
\end{abstract}

\section{Introduction} \label{sec:intro}
In the early stages of their evolution, low mass protostars launch outflows: fast flows of gas that can generate shocks when colliding with the surrounding medium. These  shocked regions often have a rich chemistry with many species experiencing significant enhancement in abundance \citep{VanDishoeck1998}. They are warm and bright, making them excellent locations to observe molecules \citep{Bachiller1997,Jorgensen2004}.\par
Further, the shocks driven by the outflow are of sufficient speed to sputter the ice mantles of dust grains, and even the dust grain material itself, into the gas phase \citep{May2000}. Thus shocked regions along protostellar outflows represent complex and interesting locations where the chemistry may be dominated by material recently released from the ices.\par
This is particularly relevant in the case of complex organic molecules (COMs). Debate over the main formation routes of these molecules \citep[eg.][]{Chuang2017,Skouteris2019} is ongoing. The most simple of these COMs, methanol (CH$_3$OH), is known to form efficiently on the grains \citep{Fuchs2009} but other more complex molecules are not well understood.\par
However, shocks present an interesting case in that it is possible that the majority of CH$_3$OH on the grains is destroyed in the sputtering process \citep{Suutarinen2014}. Thus, it is possible that some of the CH$_3$OH observed towards outflows may be formed in the post-shock gas phase, despite the efficiency of methanol formation in interstellar ices.\par
Acetaldehyde (CH$_3$CHO) has been detected in outflows \citep{Codella2015} as well as hot corinos \citep{Cazaux2003} and cold cores \citep{JimenezSerra2016}. There exist viable routes to forming the molecule on the grain surface through diffusion reactions \citep{Garrod2006} and UV processing \citep{Oberg2009} as well as in the gas phase \citep[][and references therein]{Charnley2004}.  \par
In this work, observations towards eight protostellar outflows, seven of which were found to contain many transitions of CH$_3$OH and CH$_3$CHO,  are presented. By considering the trends in the column densities of these molecules and  the ratio of CH$_3$OH isomers across different shocked regions and with different shock speeds, it is expected some information as to the origin of each molecule can be extracted.\par
Moreover, the set of outflows presented here are a sample that was initially chosen to investigate the chemical differences between NH$_3$ and H$_2$O in shocked regions \citep{Gomez-Ruiz2016}. In the outflow source L1157-B1, it was was observed that NH$_3$ transitions had a much lower terminal velocity than H$_2$O despite arising from the same source \citep{Codella2010}. Through chemical and radiative transfer modelling, this was attributed to gas phase reactions in the hot post-shock gas \citep{viti2011}. Both species were sputtered by the shock and enhanced, but in the hot, fast moving gas NH$_3$ was destroyed through reactions with H atoms. The same behaviour was observed in MHD shock models with radiative transfer \citep{Flower2012}.\par
The difference between NH$_3$ and H$_2$O were found to hold across the outflows presented here \citep{Gomez-Ruiz2016} and a search for other species that followed similar trends was attempted. In observations targeting sulfur-bearing species, a large number of CH$_3$OH and CH$_3$CHO were serendipitously detected by student researchers as part of the ORBYTS researchers in schools programme \footnote{\url{http://www.twinkle-spacemission.co.uk/orbyts/}}. In this work, the emission from those species is analysed to find the trends of their emission and abundance with shock velocity\par
In Section~\ref{sec:obs}, the observations and basic measurements are described. In Section,~\ref{sec:inference} the column density inference procedure is detailed. In Section~\ref{sec:results}, the results of these measurements are presented and in Section~\ref{sec:discuss}, the implications of the results are discussed. Finally, the work is summarized in Section~\ref{sec:conclusion}.
\section{Observations and Processing} \label{sec:obs}
Observations of eight outflow sources were obtained using the IRAM-30m telescope. The sources were chosen to match the sample observed by \citet{Gomez-Ruiz2016}. The sources, listed by central object, are presented in Table~\ref{table:sources}. Given that the offset from the central object is generally much larger than the beam width, it is assumed that the contribution from those protostars is negligible.\par
\begin{table*}
    \centering
\begin{tabular}{ccccccc}
\toprule
Source & $\alpha$ & $\delta$ & V$_{LSR}$ & L$_{bol}$ & d & Offsets Blue/Red\\
& (h:m:s) & (\si{\degree:\arcminute:\arcsecond}) & \si{\kilo\metre\per\second} & L$_{\odot}$ & pc & arcseconds\\
\midrule
NGC1333-IRAS2A & 03:28:55.4 & 31:14:35.0 & +6.0 & 25 & 235 & B(-100,+25), R(70,-15)\\
NGC1333-IRAS4A & 03:29:10.4 & 31:13:31.0 & +6.5 & 8 & 235 & B(-6,-19), R(14,25)\\
L1157 & 20:39:06.2 & 68:02:16.0 & +2.6 & 4 & 250 & B2(35,-95), R(-30,140)\\
L1448 & 03:25:38.9 & 30:44:05.0 & +4.7 & 6 & 235 & B2(-13,+29), R4(26,-125)\\
\bottomrule
\end{tabular}
    \caption{Sources and relevant parameters adapted from \citep{Gomez-Ruiz2016}.}
    \label{table:sources}
\end{table*}
The observations contain two frequency ranges, the first from \SIrange{96}{104}{\giga\hertz} and the second from \SIrange{152}{176}{\giga\hertz}. In the case of L1448-B2 and IRAS2A-B, only the second range was observed. These ranges were intended to capture transitions of many sulfur bearing species, the exploitation of which will be presented in future work.\par
Observations were carried out between the 2nd and 4th of June 2016 in wobbler switching mode. Pointing was monitored regularly, using Mercury, and found to be stable. The atmospheric opacity was low, $\tau < 0.15$.  The observations were taken using the IRAM-30m telescope's EMIR receivers and the Fourier Transform Spectrometer back end. This gave a spectral resolution of \SI{200}{\kilo\hertz} or, equivalently, velocity resolutions between \SI{0.35}{\kilo\metre\per\second} and \SI{0.6}{\kilo\metre\per\second}. All spectra were smoothed and resampled to a common velocity resolution of \SI{1.0}{\kilo\metre\per\second} for analysis. This allowed comparison between sources and greatly improved the signal to noise ratio of weak lines.\par
The data were manually inspected using the GILDAS CLASS software package\footnote{\url{http://www.iram.fr/IRAMFR/GILDAS}}. This was primarily for initial line identification. Where a peak in the spectrum was above three times the noise level, it was identified by comparing the rest frequency of the brightest channel to the JPL catalog \citep{pickett1998} accessed via Splatalogue\footnote{\url{http://www.splatalogue.net/}}. Transitions of the E isomer of CH$_3$CHO and both the A and E isomers of CH$_3$OH were detected in every source except L1448-R4. The L1448-R4 spectra were badly affected by absorption at the systematic velocity of L1448 and so this source is not discussed further.\par
Whilst many transitions of the species were detected, many more transitions are present in the observed frequency ranges. In order to make use of these non-detections, the CLASS WEEDS library and the JPL catalog were used to extract every transition in the observed frequency range. However, this was a prohibitively large number of transitions and so limits on $E_u$ and $A_{ij}$ were chosen based on the properties of the manually detected lines. Only transitions within these limits were exported for further processing. This included every CH$_3$OH transition with an $E_u < \SI{250}{\kelvin}$ and $A_{ij} > \SI{e-6}{\per\second}$ and every CH$_3$CHO transition with an E$_u < \SI{100}{\kelvin}$ and $A_{ij} > \SI{e-5}{\per\second}$. These ranges include every detected transition in the observed frequency range but limited the number of non-detections in the data set. Spectra covering a velocity range of \SI{200}{\kilo\metre\per\second} centred on each transition were exported regardless of whether a peak had been identified in the spectrum. \par
The exported spectra were baseline subtracted in python by fitting polynomials of up to order 3 to the spectra using numpy's polyfit. The rms difference between the baseline and the spectra was then calculated to give the spectrum rms. The smoothed spectra have rms values between 3 and \SI{10}{\milli\kelvin}. Transitions were classed as detections if the peak emission was above three times the rms noise level. The spectra of detected lines were median stacked and the velocities at which the median stacked peak fell to the rms emission level were recorded. These were considered to be the minimum and maximum velocity limits of the emission of that species in that source. All transitions of the species were then integrated between these limits even if a peak was not detected. The uncertainty on the integrated emission of each transition was found by propagating the uncertainties due to the noise and velocity resolution of their spectra as well as a 20\% calibration error. \par
Finally, transitions where the integrated emission was larger than the noise level were inspected visually. Any transition which was blended with the emission of another species was removed from the data set. Two groups of three CH$_3$OH transitions were blended in every source. For these, the distance between the peaks of the blended transitions in velocity space was added to the integration limits such that all blended lines were integrated together. The total flux of these lines was treated specially as detailed in Section~\ref{sec:ch3ohflux}.\par
In the online appendix, the measured properties of every transition in the observed frequency ranges are presented in tables for each source in Appendix~\ref{app:tables}. In Appendix~\ref{app:spectra}, plots of the spectra of every detected line of each species in each source are also presented.\par
\section{Inference of Gas Properties} \label{sec:inference}
The integrated emission of the CH$_3$OH and CH$_3$CHO transitions were used to infer the column density of each species towards each source as well as the gas properties through model fitting. Since any model that predicts strong emission for an undetected transition cannot be correct, the measured emission of every line of each species was included in the fits regardless of whether or not a peak was detected. In this section, the model used for each species and the fitting procedure is detailed.\par
\subsection{CH$_3$OH Fluxes}
\label{sec:ch3ohflux}
The LAMDA database\footnote{\url{http://home.strw.leidenuniv.nl/~moldata/CH3OH.html}} contains collisional data for the two isomers of CH$_3$OH (A-CH$_3$OH and E-CH$_3$OH) orginally computed by \citet{Rabli2010}. This allowed the use of RADEX \citep{VanderTak2007} to predict the emission of each transition. RADEX requires the column density of each isomer, the gas temperature, and the linewidth as input parameters. \par
RADEX assumes the source fills the telescope beam and so the fluxes had to be adjusted for a fifth free parameter, the source size. To do this the source size was used to calculate a filling factor,
\begin{equation}
    \eta_{ff}=\frac{\theta_S^2}{\theta_S^2+\theta_{MB}^2}
\end{equation}
where $\theta_S$ is the source size and $\theta_{MB}$ is the beamsize which is a function of frequency. The RADEX outputs were multiplied by this factor to account for the beam dilution present for an assumed source size.\par
Fitting the integrated emission directly provided a convenient solution to the blended transitions of CH$_3$OH. The total integrated area of each set of blended transitions was measured from the spectra and compared to the sum of the RADEX emission for those transitions. If a RADEX model fits the individual transitions in a data set, it follows that it would fit the sum of any set of transitions. Therefore, fitting the sum of blended transitions allows them to contribute information to the fit, avoiding any approximations that are often used to separate the flux from blended transitions such as assuming LTE. However, it does create a degeneracy in the fits as the relative emission of each line is not considered. Where many other transitions are detected, this degeneracy is resolved. If few other transitions are detected, the reported uncertainties are significantly larger.
\subsection{CH$_3$CHO Fluxes}
For CH$_3$CHO, no collisional data was available and so the integrated emission was predicted by assuming optically thin emission and LTE conditions. Assuming every transition has an upper state column density ($N_u$) given by a Boltzmann distribution, the upper state column density can be related to the integrated emission ($F_T$) and rearranged to give that emission as a function of the column density ($N$),  rotational temperature ($T_{rot}$), and the source size or filling factor,
\begin{equation}
    F_t= \frac{hc^3}{8\pi k_B} \frac{A_{ij}g_u}{\nu^2}\eta_{ff}\frac{N}{Z}\exp{\left(\frac{-E_u}{k_B T}\right)}
\end{equation}
The result is an approach that is effectively equivalent to the rotation diagram method \citep{goldsmith1999} but allows the source size to be accounted for in a more direct fashion and is more similar to the method used for CH$_3$OH.\par
\subsection{Inference Procedure}
The likelihood of having obtained the data given a set of parameters can be defined as,
\begin{equation}
P(\boldsymbol{F_d} | \boldsymbol{\theta})=\exp\left(-\frac{1}{2}\sum_i\frac{(F_{d,i}-F_{t,i})^2}{\sigma_{F,i}^2}\right)
\end{equation}
where $F_{d,i}$ and $F_{t,i}$ are the observed and model integrated emission of transition $i$, $\sigma_F$ is the observational uncertainty, and $\boldsymbol{\theta}$ is the set of parameters which may include the gas temperature and column density. From here, Bayes' theorem can be applied to obtain the probability distribution of the parameters given the observed data,\par
\begin{equation}
    P(\boldsymbol{\theta}|\boldsymbol{d})=\frac{P(\boldsymbol{d} | \boldsymbol{\theta})P(\boldsymbol{\theta})}{P(\boldsymbol{d})}
    \label{eq:bayes}
\end{equation}
where $P(\boldsymbol{\theta})$ is the prior probability of the parameters and $P(\boldsymbol{d})$ is the Bayesian evidence, which can simply be treated as a normalizing factor.\par
An advantage of this method is that poorly constrained variables can be marginalized over. For example, the angular size of each source is an unknown in this work. However, if it is treated as a free parameter, the uncertainty this introduces to the model will be reflected in the marginalized posterior distributions of the other parameters.\par
The posterior distribution was sampled using the python package emcee \citep{Foreman-Mackey2013}. This is an affine-invariant sampler \citep{Goodman2010} in which multiple chains of samples are created by ``walkers'' which each sample the space. Each walker moves through the parameter space generating a chain of parameter samples, starting with a random value of the parameters of interest drawn from the prior distributions. However, unlike many MCMC samplers, walkers do not move randomly and are instead informed by each other.\par
The prior probability distributions were simply assumed to be flat between limits. The temperature prior was non-zero between 0 and \SI{150}{\kelvin} and the source size prior was non-zero between 0 and \SI{200}{\arcsecond}. The column density prior was non-zero between \num{e11} and \SI{e17}{\per\centi\metre\squared} and, in the case of CH$_3$OH, the H$_2$ density prior was non-zero between \num{e3} and \SI{e8}{\per\centi\metre\cubed}. The column density and gas density were sampled in log-space for numerical reasons as large values will otherwise dominate the sampling.\par
For the parameter inference, 12 walkers were used and the chains were run for \num{e5} steps per walker. The Geweke diagnostic \citep{Geweke1992} was used to verify that the chains had converged.\par
\section{Results}\label{sec:results}
The gas densities and kinetic temperature of the gas traced by CH$_3$OH as well as the column densities of both species in each object obtained from the parameter inference are given in Tables~\ref{table:summary-ch3oh} and \ref{table:summary-ch3cho}. Most likely values are presented for the column densities as well as the 67\% probability intervals of all parameters. This implies that if the model is correct, there is 67\% chance the true value of each parameter lies within the given range. \par
These values should be treated with some caution as RADEX gives an imperfect representation of a slab of gas described by a single density and temperature and this not the ideal model for a shocked region. This is even more of an issue for the LTE results presented for CH$_3$CHO. Further, in the best case, the parameter values obtained will represent the average values of the gas encompassed by the telescope beam. However, in both cases, the model results do give some insight and represent the best models available to us. All of the results of the fitting are discussed below.
\subsection{CH$_3$OH - Gas Properties}
In Table~\ref{table:summary-ch3oh}, third and fourth columns give the 67\% probability range of the gas density and temperature. In the weakest emitting sources, the kinetic temperature is poorly constrained and only an upper limit is reported. This is due to the low range of E$_u$ values of the detected transitions. The temperature only needs to be sufficiently low not to excite the higher energy undetected transitions and beyond this, there is no change in the fit. \par
In sources with a large number of detected lines, the gas temperatures are well constrained. There is no clear variation between these sources, most giving values between 40 and \SI{50}{\kelvin}. Whilst this is much lower than the maximum temperature typically reached in the shocks used to model outflows \citep{Flower2015}, a single temperature and gas density is being fit to a shock structure that will encompass a wide range of gas conditions. Further, it is consistent with that found in other protostellar shock sources \citep[eg. CO in L1157-B1][]{lefloch2012}, including that found previously for CH$_3$OH in L1157-B2 \citep{Mcguire2015}. \par
\begin{table*}
\centering
\begin{tabular}{c|ccc|cccc}
\toprule
Source & Lines & T$_{Kin}$ & n$_{H2}$ & N(A-CH$_3$OH) & N(A-CH$_3$OH) Range & N(E-CH$_3$OH) & N(E-CH$_3$OH) Range \\
 & & K & \ $\times$ \SI{e5}{\per\centi\metre\cubed} & \si{\per\centi\metre\squared} & \si{\per\centi\metre\squared} & \si{\per\centi\metre\squared} & \si{\per\centi\metre\squared} \\
\midrule
IRAS2A-R & 28 & 49.1 - 57.8 & 3.8 - 5.3 & \num{9.8e+14} & \num{7.8e+14} - \num{1.3e+15} & \num{5.2e+14} & \num{4.7e+14} - \num{5.9e+14} \\
IRAS2A-B & 7 & \textless~41.7 & 2.1 - 27.6 & - &  \textless~\num{2.23e+15} & \num{7.3e+13} & \num{6.7e+13} - \num{2.3e+14} \\
IRAS4A-R & 21 & 38.0 - 45.6 & 4.9 - 6.1 & \num{1.9e+14} & \num{1.5e+14} - \num{2.4e+14} & \num{1.1e+14} & \num{9.6e+13} - \num{1.2e+14} \\
IRAS4A-B & 25 & 40.7 - 46.4 & 4.9 - 6.0 & \num{6.9e+14} & \num{5.8e+14} - \num{8.0e+14} & \num{3.1e+14} & \num{2.8e+14} - \num{3.4e+14} \\
L1157-R & 3 & \textless 136.5 & 1.7 - 5.0 & \num{3.9e+13} & \num{2.7e+13} - \num{6.7e+13} & \num{3.7e+12} & \num{2.0e+12} - \num{8.2e+12} \\
L1157-B2 & 25 & 41.9 - 47.1 & 2.8 - 4.1 & \num{9.1e+14} & \num{7.7e+14} - \num{1.1e+15} & \num{3.8e+14} & \num{3.3e+14} - \num{4.3e+14} \\
L1448-B2 & 1 & \textless~22.9 & 1.6 - 8.0 & - &  \textless~\num{1.54e+14} & \num{4.4e+12} & \num{3.0e+12} - \num{8.8e+12} \\
\bottomrule
\end{tabular}
\caption{Summary of the RADEX fits to CH$_3$OH. The first column gives the number of lines with peaks above three times the noise level, the second and third are the 67\% probability intervals for the gas temperature and density. The remaining columns are the best fit column density and 67\% probability interval for each isomer of CH$_3$OH. 3 sigma upper limits are given where no lines were detected.}
\label{table:summary-ch3oh}
\end{table*}
The gas densities do not vary greatly, only the sources associated with IRAS4A have most likely ranges that are differ significantly from $\sim$\SI{4e5}{\per\centi\metre\cubed}. However, it is interesting to note that each pair of sources associated with the same central objects are very similar. Whilst the IRAS2A-B gas density posterior distribution is quite wide, the majority of the probability density is in the same range as that reported for IRAS2A-R. Further, the IRAS4A and L1157 sources are remarkably consistent.\par
\subsection{CH$_3$OH - Column Density}
\begin{figure*}
\centering
\includegraphics[width=\textwidth]{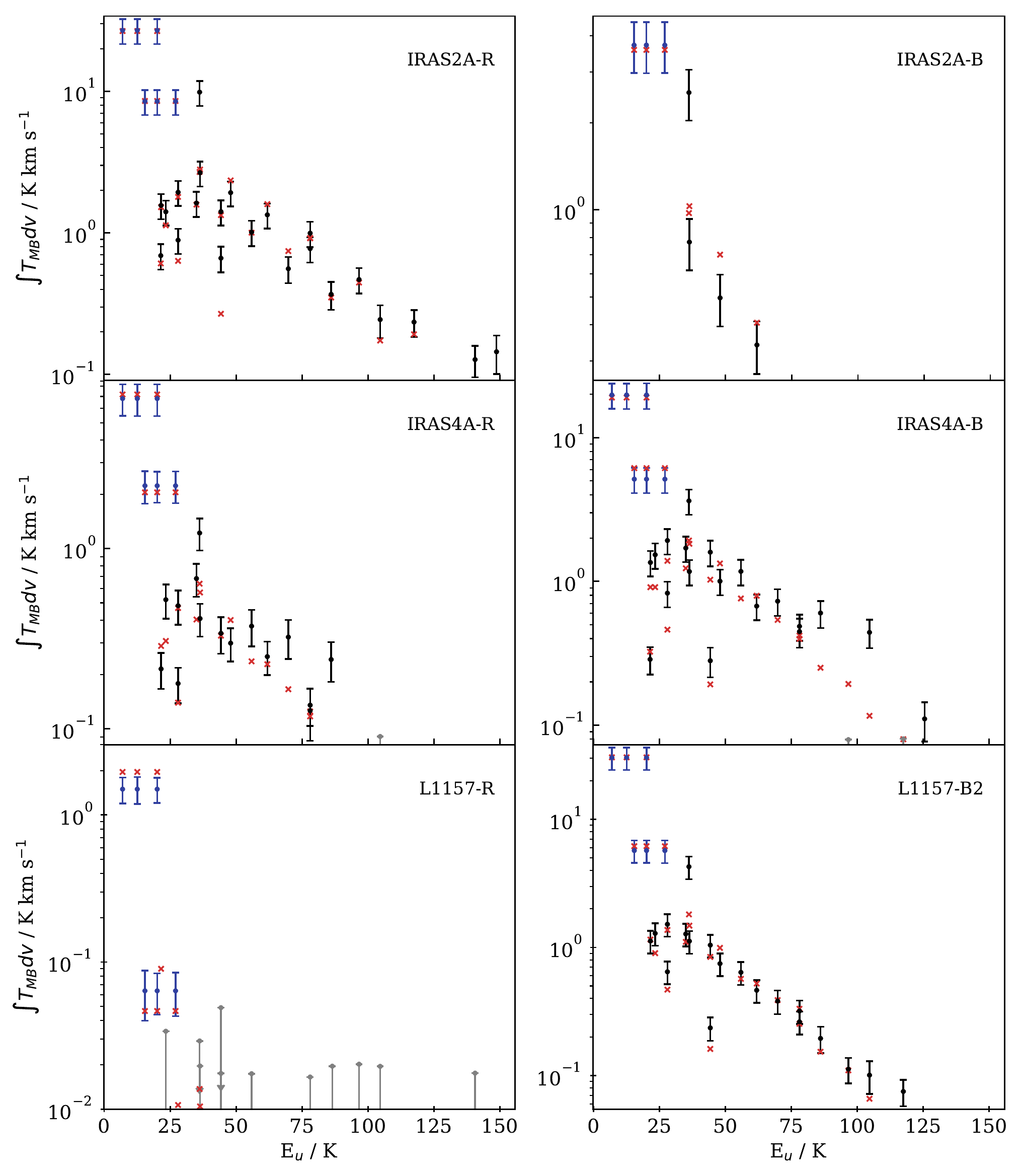}[h]
\caption{Measured flux (black points and error bars) of CH$_3$OH transitions plotted against E$_u$. The blue points are those transitions that are blended, the total emission across all three transitions is plotted in each case along with the sum of the RADEX model emission. The grey upper limits show undetected transitions. The red crosses indicate the fluxes produced by RADEX for the most likely parameter values.}
\label{fig:flux-ch3oh}
\end{figure*}

In Figure~\ref{fig:flux-ch3oh}, the measured flux of each observed CH$_3$OH transition is plotted. In red, the fluxes obtained from RADEX using the most likely values of the parameters are also plotted. The agreement is good considering the use of a single gas component to fit a shocked region, with $\chi^2$ values of $\sim$1.5. In Table~\ref{table:summary-ch3oh}, the most likely column density of each CH$_3$OH isomer is given along with the 67\% confidence interval. In the case of L1448-B2 and IRAS2A-B, only an upper limit can be given for A-CH$_3$OH. \par
CH$_3$OH has been previously detected towards each of these sources. The most likely column densities from this work were compared to the column density of CH$_3$OH found towards the outflows of IRAS-2A \citep{Jorgensen2004}, IRAS-4A \citep{Blake1995}, L1448 \citep{Tafalla2010}, and L1157 \citep{Burkhardt2016}. The observations in those works covered fewer lines so only total CH$_3$OH column densities were reported. However, for every source, the sum of the isomer column densities obtained through this analysis is consistent with the value reported in those works. \par
\subsection{CH$_3$CHO Column Density}
The results of the LTE parameter inference for CH$_3$CHO emission from the observed sources are summarized in Table~\ref{table:summary-ch3cho}. No transitions of CH$_3$CHO were detected at a significant level towards IRAS2A-B, L1157-R and L1448-B2. As a result, the posterior distribution of N only has upper bounds in these sources and only the upper limit is reported.\par
\begin{figure*}\centering
\includegraphics[width=0.9\textwidth]{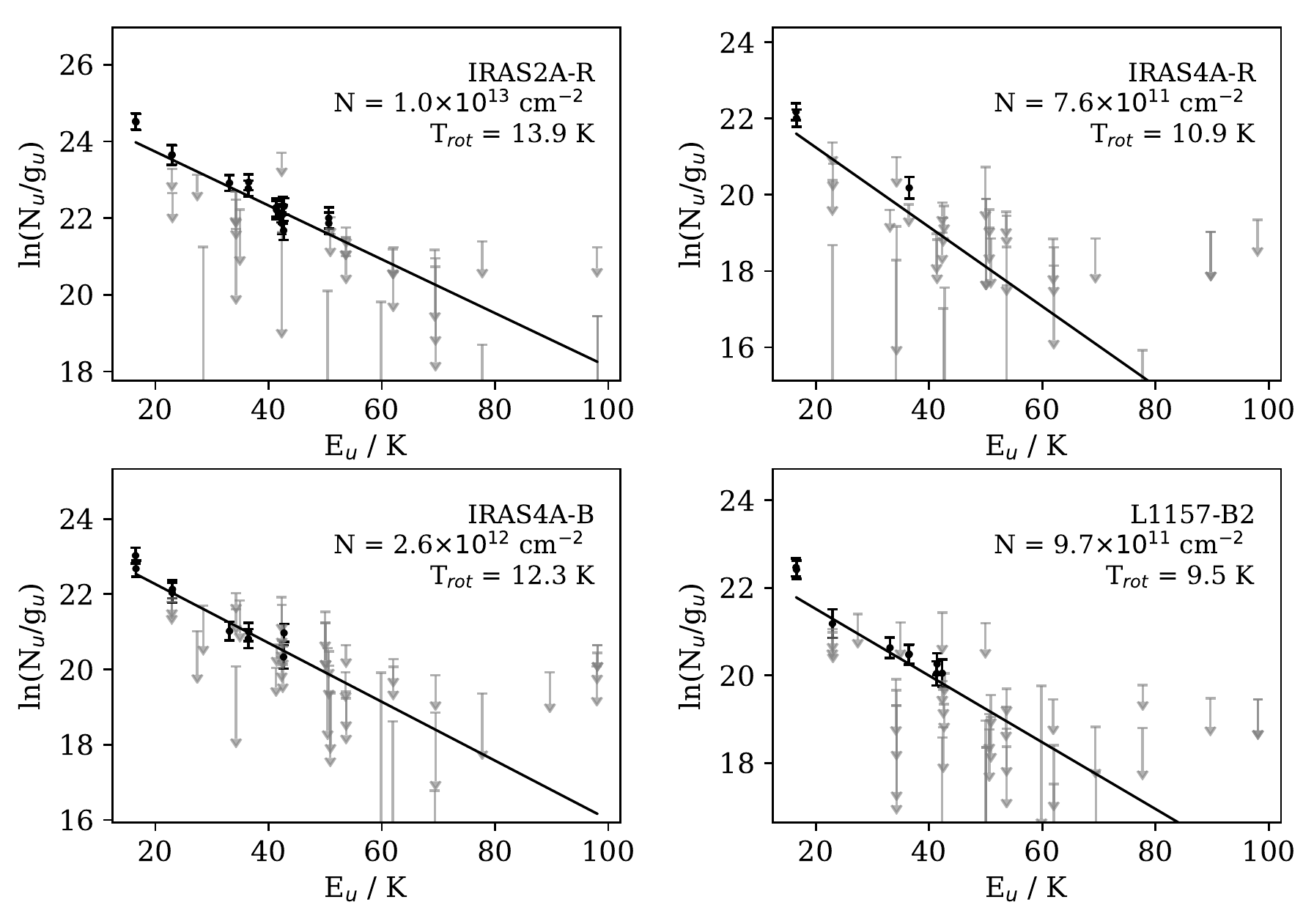}
\caption{Rotation Diagrams of CH$_3$CHO in each source. The black line shows the Boltzmann distribution given by the most likely values of $N$ and $T$ for each source. Black points are detected transitions adjusted for the best fit filling factor and grey points are the upper limits calculated from the non-detections.}
\label{fig:rot-ch3cho}
\end{figure*}

\begin{table*}
\centering
\begin{tabular}{c|cccc|c}
\toprule
Source & Lines & T$_{rot}$ & N & N(CH$_3$CHO) Range & CH$_3$CHO/CH$_3$OH\\
 & & K & \si{\per\centi\metre\squared} & \si{\per\centi\metre\squared} \\
\midrule
IRAS2A-R & 17 & 12.6 - 15.5 & \num{1.0e+13} & \num{5.8e+12} - \num{2.4e+13} & \num{6.7e-03} \\
IRAS2A-B & 0 & - & - &  \textless \num{1.44e+12} & \textless \num{4.3e-03} \\
IRAS4A-R & 3 & 6.1 - 15.2 & \num{7.6e+11} & \num{5.5e+11} - \num{1.0e+12} & \num{2.6e-03} \\
IRAS4A-B & 9 & 11.1 - 13.7 & \num{2.6e+12} & \num{2.1e+12} - \num{3.2e+12} & \num{2.6e-03} \\
L1157-R & 0 & - & - &  \textless \num{7.47e+11} & \textless \num{5.2e-03} \\
L1157-B2 & 9 & 8.4 - 12.1 & \num{9.7e+11} & \num{8.1e+11} - \num{6.3e+12} & \num{7.6e-04} \\
L1448-B2 & 0 & - & - &  \textless \num{3.40e+11} & \textless \num{3.8e-03} \\
\bottomrule
\end{tabular}
\caption{Summary of the LTE fits to CH$_3$CHO. The first column gives the number of lines with peaks above three times the noise level, the second is the best fit rotation temperature and the third and fourth columns are the best fit column density and 67\% probability interval. Where only an upper limit on the column density was found, the three sigma upper limit is given. The final column is the ratio of the CH$_3$CHO column density to the CH$_3$OH column density measured in the source.}
\label{table:summary-ch3cho}
\end{table*}
The fits to the CH$_3$CHO fluxes are presented as rotation diagrams in Figure~\ref{fig:rot-ch3cho}. The best fit $N$ and $T$ are used to plot the expected Boltzmann distribution as a straight line. The points are the upper state column density of each transition, calculated from the measured integrated emission and adjusted for the filling factor using the most likely value of $\theta_S$. \par
\subsection{"Nuisance" Parameters}
There are two parameters that are not of particular interest to the analysis but are important for the fitting, these are the line width in the case of the RADEX fits and the filling factor for both the RADEX and LTE fits.\par
\begin{figure*}
\centering
\includegraphics[width=\textwidth]{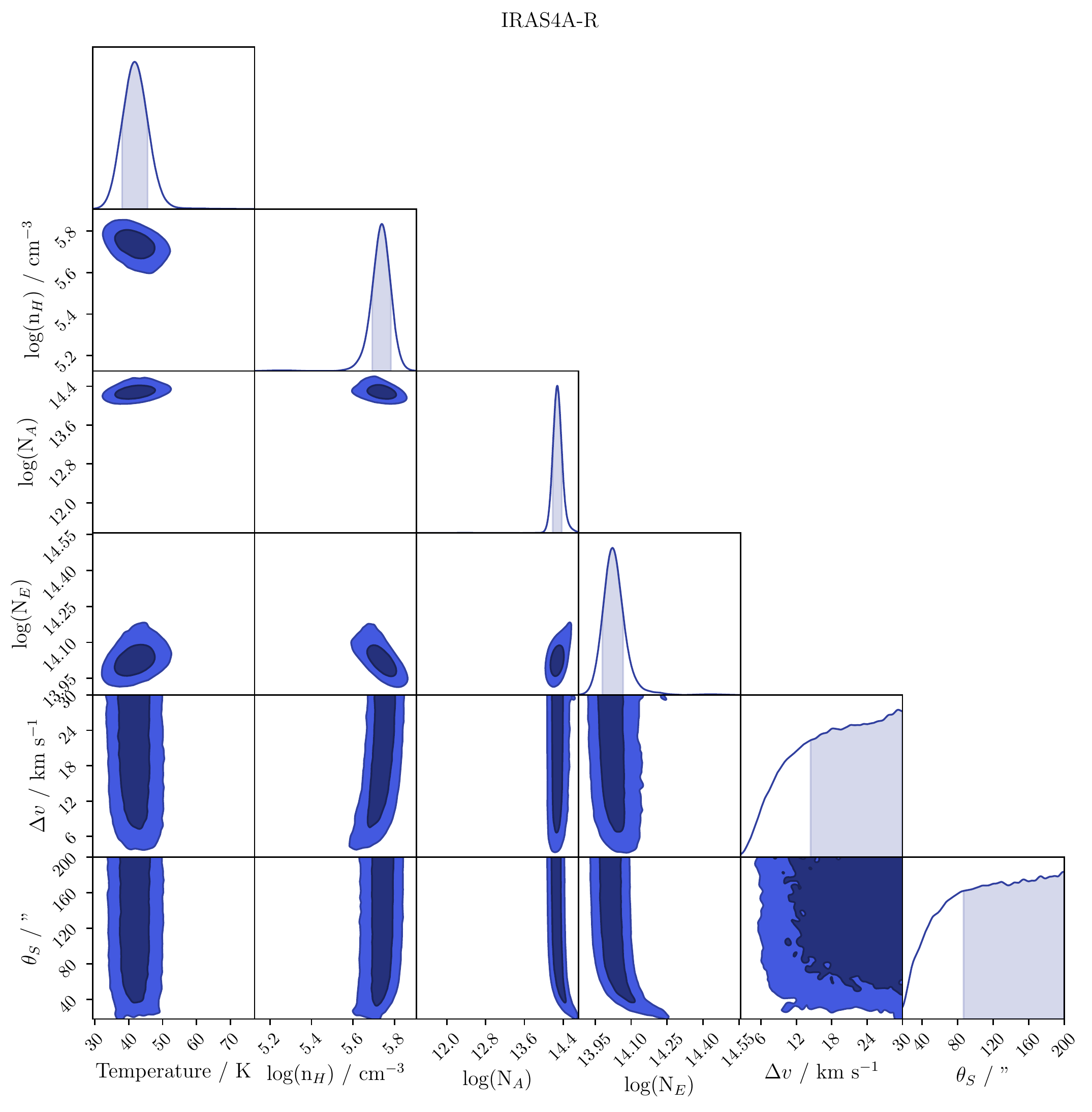}
\caption{Joint and marginalized posterior distributions of the fitted parameters. In the joint plots, the blue contours show the 67\% and 95\% probability intervals. In the marginalized plots, the shaded area shows the 67\% interval. The source size tends to large values indicating the best fits are found assuming source fills the beam.}
\label{fig:corner}
\end{figure*}

In Figure~\ref{fig:corner}, it can be seen that when the linewidth is allowed to vary freely, the probability distribution tends to large values. From the RADEX manual \footnote{\url{http://home.strw.leidenuniv.nl/~moldata/radex_manual.pdf}}, one finds that the manner in which the flux and line intensity are calculated means that the line width is irrelevant in the limit of low optical depth. Thus the fits are tending to a regime that is consistent with optically thin CH$_3$OH emission.\par
The line width is only constrained in the case of IRAS2A-R, in which the most likely value is half the full width of the detected lines. A second parameter inference was run, fixing the line widths to their FWHM values and consistent results were found for the other parameters. Thus it is considered that the linewidth is unimportant and is left as a free parameter as this increases the reported uncertainties and is thus more conservative. This is appropriate because the lines are not Gaussian and RADEX assumes they are.\par
Figure~\ref{fig:corner} also shows a common result for the source size: it also tends to large values. In these cases, it is reasonable to assume the source fills the beam. For source sizes much larger than the beam, the filling factor tends to unity so the actual value becomes irrelevant and thus any source that fills the beam will have a posterior that simply tends to high values rather than a specific value.  From the figure, it can be seen the column densities and the source size are correlated below $\sim$80". This broadens the column density distributions and is thus included in the reported column density uncertainty.\par
In other sources, the filling factor probability distribution is close to constant across transitions for many source sizes. Most transitions of CH$_3$OH in the observed frequency range are associated with beam sizes of 15" with only a small number of transitions reaching 25" thus, in a source with few detections the source size is not constrained. This manifests as a broader variation of column densities as the filling factor is a constant offset and a low source size can be compensated by a smaller column density and vice versa.\par
\section{discussion}\label{sec:discuss}
\subsection{E/A Methanol Ratio}
\label{sec:isomers}
Methanol has two spin isomers referred to as E and A-CH$_3$OH respectively. Since the conversion of one isomer to another is a strongly forbidden process, it is often assumed in that this ratio is fixed at formation and then preserved \citep{Kawakita2009}.\par
Whilst this assumption is typically applied to methanol observed in cometary ices, \citet{Wirstrom2011} argue this should also be true for the interstellar medium. They argue that the timescale for the reactions that would alter the ratio are long compared to the lifetime of an average methanol molecule and thus conversion must be negligible.
The ratio of the E and A isomers of CH$_3$OH can be used to calculate a characteristic temperature ($T_{spin}$) through the equation,
\begin{equation}
    \frac{E}{A}=\frac{\sum_i g_i\exp\left(\frac{-E_i}{kT_{spin}}\right)}{\sum_j g_j\exp\left(\frac{-E_j}{kT_{spin}}\right)}
\end{equation}
where the sums over $i$ and $j$ are sums over all transitions of E and A-CH$_3$OH respectively. If the methanol was thermalized at formation, $T_{spin}$ represents the temperature of the formation environment.\par
\begin{figure}
\includegraphics[width=0.5\textwidth]{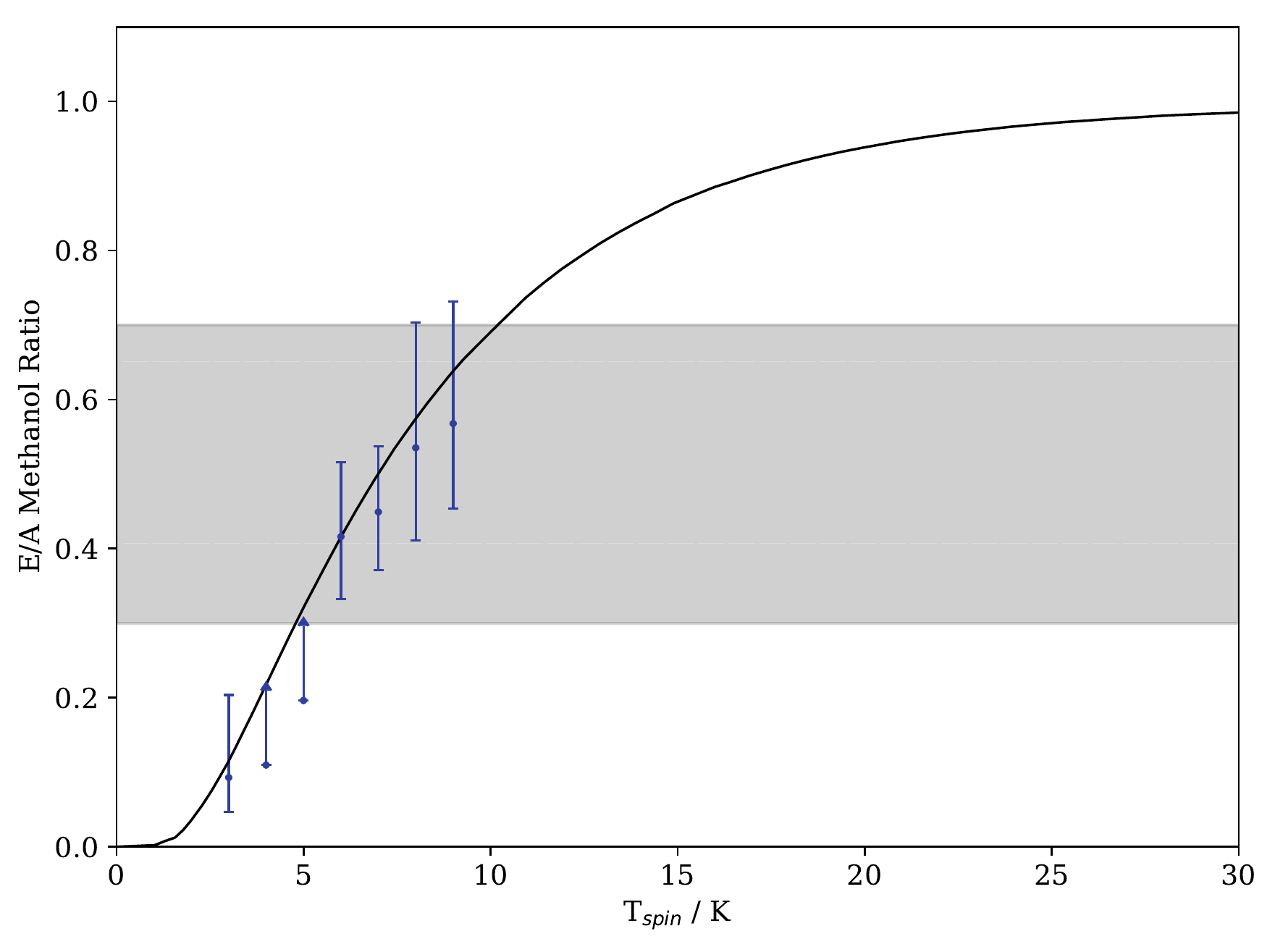}
\caption{Ratio of E/A methanol column densities towards each source plotted as points with errorbars. Black curve indicates the expected value of the ratio for methanol formed at a given temperature. Grey band shows the range of ratios expected for methanol formed between 5 and \SI{10}{\kelvin}. Lower limits are those sources for which only an upper limit on A isomer of methanol were found.}
\label{fig:spintemps}
\end{figure}

Assuming this is the case, the ratio of E to A-CH$_3$OH in the observed sources can be calculated to give the spin temperature and hence the temperature at which the methanol in those sources formed. Figure~\ref{fig:spintemps} shows that the ratio in every source. Most sources are consistent with a spin temperature between 5 and \SI{10}{\kelvin}. The one source which has an unusually low ratio but is not an upper limit is L1157-R. Only one A-CH$_3$OH transition was detected in this source and it was part of the blended triplet at $\sim$\SI{96}{\giga\hertz}.\par
Such low and similar spin temperatures are inconsistent with a scenario in which methanol forms in the post-shock gas in the outflow. If thermalization is assumed, the spin temperature would equal the post-shock gas temperature, which is $\sim$\SI{50}{\kelvin} in each source. Thus it is concluded that that the methanol observed in these outflows formed before the shock on the grain surfaces before being released by shock sputtering. This is corroborated by \citet{Fuchs2009} who showed CO efficiently hydrogenates to form methanol between 12 and \SI{20}{\kelvin} on interstellar ice analogues.
\subsection{Velocity Trends of CH$_3$OH}
\label{sec:source-comp}
\begin{figure}
\includegraphics[width=0.5\textwidth]{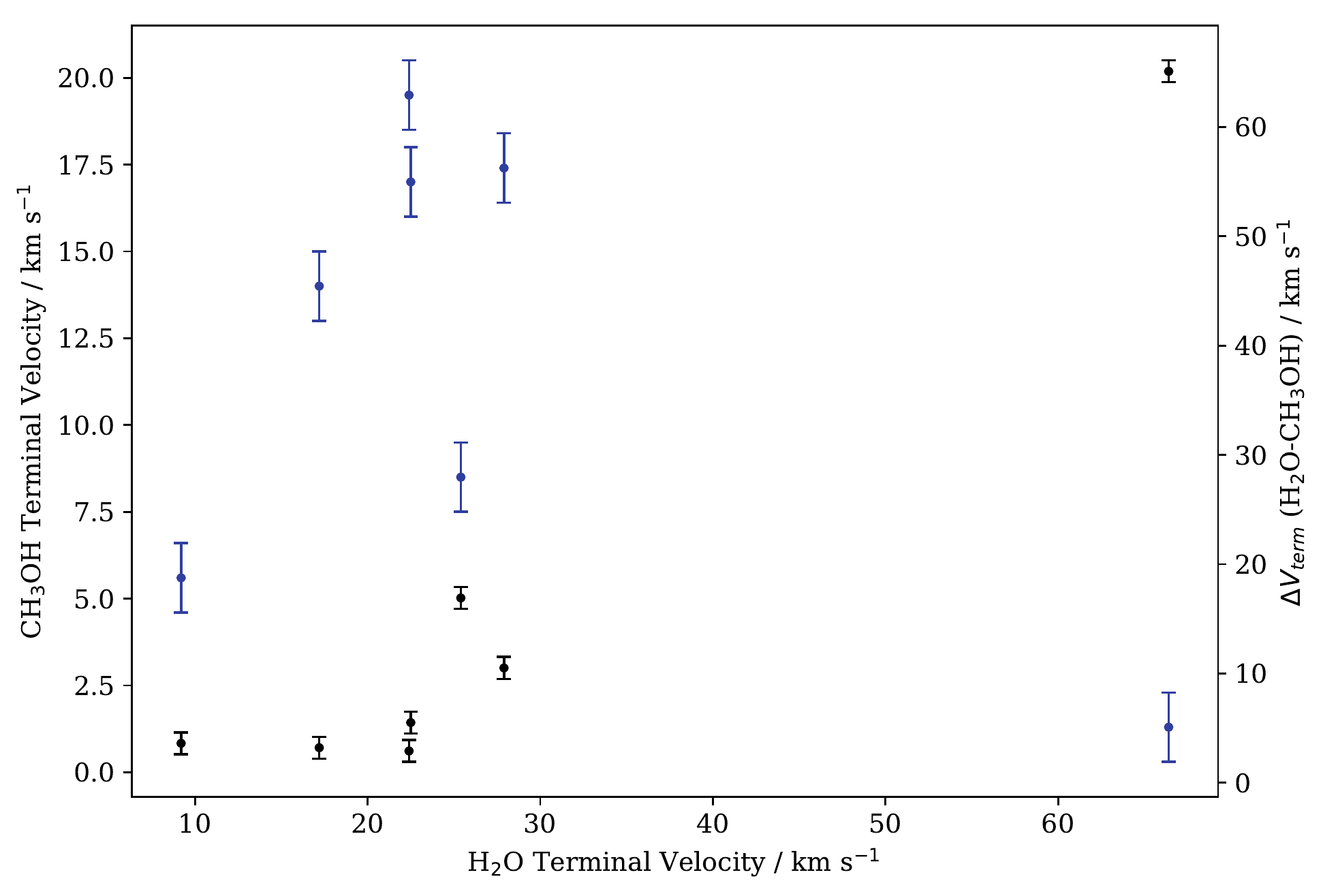}
\caption{Terminal velocity of CH$_3$OH against terminal velocity of H$_2$O used as a proxy for the outflow velocity. The blue points show the terminal velocity (left axis) and the black show the difference between the two species' terminal velocities (right axis).}
\label{fig:ch3oh-velocity}
\end{figure}

In section~\ref{sec:intro}, the difference in terminal velocities between NH$_3$ and H$_2$O in the observed sources is discussed. The terminal velocity of H$_2$O in these sources was taken from \citet{Gomez-Ruiz2016} and compared to CH$_3$OH. The terminal velocity is the velocity at which the emission peak reaches the rms level and is given as an absolute value relative to the local standard of rest. Figure~\ref{fig:ch3oh-velocity} shows that CH$_3$OH follows the same trend as NH$_3$ \citep[Figure 8,][]{Gomez-Ruiz2016}. Up to a H$_2$O terminal velocity of \SI{30}{\kilo\metre\per\second}, the CH$_3$OH terminal velocity increases with the H$_2$O velocity. However, the difference between the H$_2$O and CH$_3$OH terminal velocity increases steadily and at much higher velocities, the CH$_3$OH terminal velocity decreases drastically. This is consistent with previous observations of H$_2$O and CH$_3$OH transitions towards IRAS2A and IRAS4A \citep{Suutarinen2014}.\par
The velocity trend is consistent with CH$_3$OH behaving similarly to NH$_3$. That is, it is released from the ice mantles by the sputtering in the shock and is then destroyed in the post-shock gas phase to a degree that is dependent on the shock speed. This is corroborated by the results of Section~\ref{sec:isomers}, which showed that the isomer ratio of methanol is consistent with a grain surface origin. It is also the case that the fractional abundance of CH$_3$OH is lower in sources with higher H$_2$O terminal velocities. \par
There are no gas phase destruction routes for CH$_3$OH in the KIDA or UMIST databases that have large energetic barriers. However, a reaction with H with an energetic barrier of $\sim$\SI{3000}{\kelvin} exists in the atmospheric chemistry literature \citep{Sander2011ChemicalEvaluation}. Alternatively, as suggested by \citet{Suutarinen2014}, the sputtering process could destroy the CH$_3$OH molecules if the shock is sufficiently high speed. \citet{Suutarinen2014} find speeds $>$\SI{25}{\kilo\metre\per\second} are required to observe the destruction of methanol which is confirmed here as the first major drop in CH$_3$OH abundance occurs in L1157-R which has  a H$_2$O terminal velocity of \SI{27}{\kilo\metre\per\second}.
\subsection{Velocity Trends of CH$_3$CHO}
CH$_3$CHO also shows an extremely similar terminal velocity trend to that seen in figure~\ref{fig:ch3oh-velocity}. This implies that CH$_3$CHO is also destroyed through a process that is more efficient in faster shocks.\par
\begin{figure}
\includegraphics[width=0.45\textwidth]{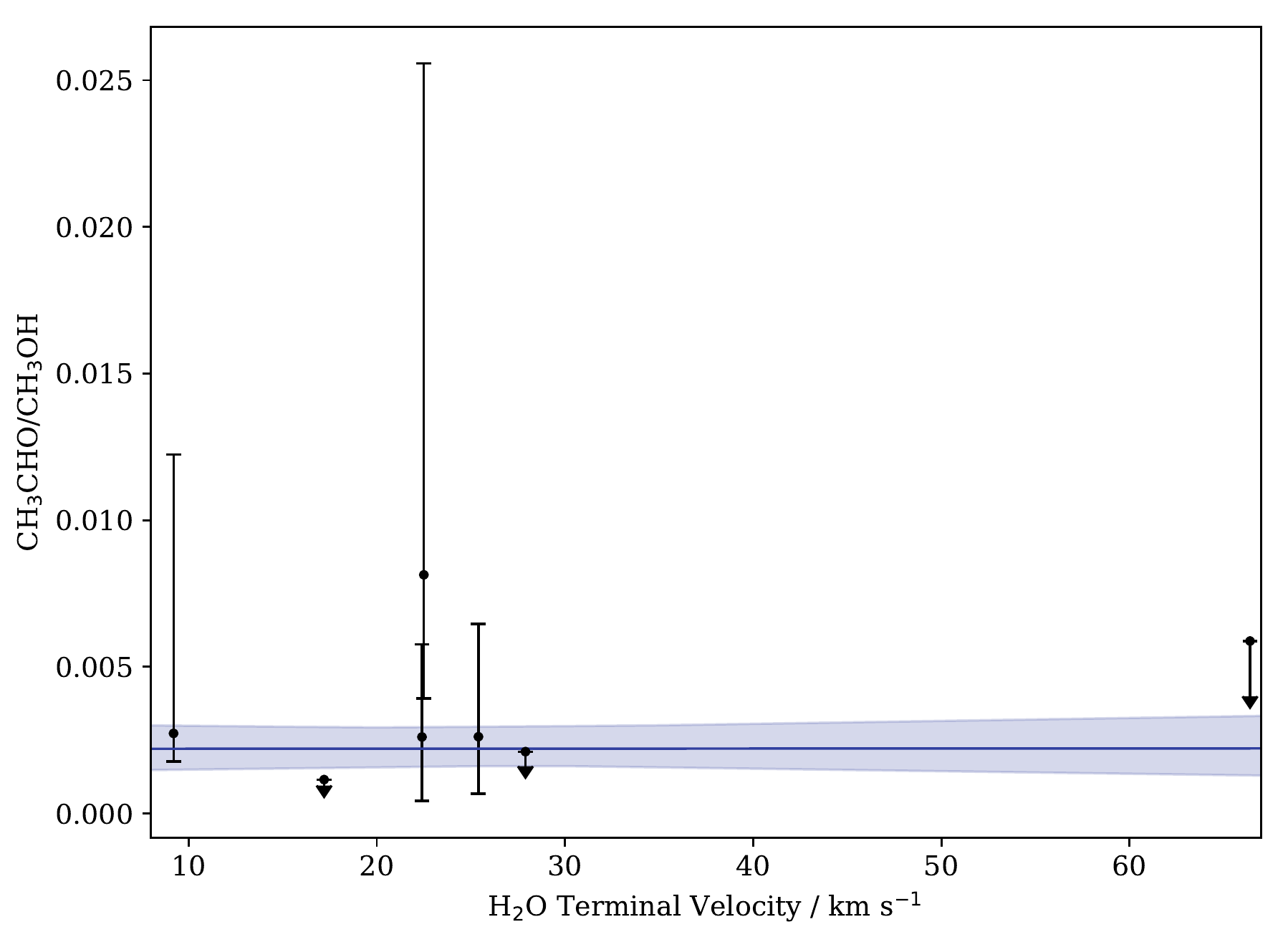}
\caption{Ratio of CH$_3$CHO and CH$_3$OH column densities against a proxy for the outflow velocity. The ratio appears to is no evidence for variation of the ratio with velocity.}
\label{fig:ch3cho-ch3oh-ratio}
\end{figure}

Figure~\ref{fig:ch3cho-ch3oh-ratio} shows the ratio of the column densities of CH$_3$CHO and e-CH$_3$OH in each source. By drawing random column densities from the MCMC chains, probability distributions for the ratios can be calculated and their most likely values and 67\% probability intervals are plotted as error bars. From the probability distributions, the relationship between the ratios and the H$_2$O terminal velocity can also be found. The most likely case is that the ratio is independent of the shock velocity and is constant at 0.002. This constant ratio and the variation due to uncertainty in the fit are plotted as a blue line and shaded region respectively.  \par
Taken together, these two facts indicate a grain surface origin of CH$_3$CHO. The velocity behaviour shows that in the post-shock gas of fast outflows, CH$_3$CHO is largely destroyed in the gas phase. Whereas the approximately constant ratio of CH$_3$CHO and CH$_3$OH in lower velocity outflows is inconsistent with CH$_3$OH being formed in the pre-shock grains and CH$_3$CHO being formed in post-shock gas. The initial abundance of CH$_3$OH should be independent of the shock properties (assuming a fast enough shock for complete sputtering of the ice mantles) but the amount of gas phase synthesis of CH$_3$CHO would be strongly tied to the conditions of the post-shock gas before it reaches temperatures that destroy CH$_3$CHO.\par
\section{Conclusions} \label{sec:conclusion}
Eight outflow sources have been observed and, in seven, transitions of CH$_3$OH and CH$_3$CHO were detected. These have been analysed to determine the origin of those species in outflows and their post-shock behaviour.\par
The integrated emission of a large set of transitions of each species in the observed frequency range were measured and used to infer the column density of each species as well as the gas density and temperature fits from RADEX for CH$_3$OH. All of these quantities are summarized in Tables~\ref{table:summary-ch3oh} and \ref{table:summary-ch3cho}. Whilst these are informative, shock structures are likely to contain regions with highly varying gas properties. The higher resolution view provided by interferometry such as that being performed for CH$_3$OH by the SOLIS project (e.g. De Simone et al in prep.) would greatly improve our understanding of such objects.\par
From column density ratios of CH$_3$OH isomers, it is concluded that the most likely case is that CH$_3$OH is formed on the pre-shock ice mantles and released into the gas phase via sputtering. This assumes the spin temperature of the isomers was set at the formation of the molecules and reflects the temperature of the formation environment. Spin temperature of 5 to \SI{10}{\kelvin} are considered inconsistent with formation in shocked gas.\par
From the terminal velocity trends of the CH$_3$OH transitions and the CH$_3$OH fractional abundance as a function of H$_2$O, it is concluded that in sufficiently fast shocks CH$_3$OH is destroyed either during the sputtering process or by chemistry in the post shock gas phase.\par
There is no direct evidence for the formation route of CH$_3$CHO. However, the near constant ratio of CH$_3$CHO to CH$_3$OH casts doubt on a gas phase origin. The amount of gas phase production of CH$_3$CHO would vary from source to source depending on the gas properties which makes a constant ratio unlikely if CH$_3$OH is formed on dust grains.
\acknowledgements
JH, SV and JCR acknowledge funding from the STFC grant ST/M001334/1. CC acknowledges the project PRIN-INAF 2016 The Cradle of Life - GENESIS-SKA (General Conditions in Early Planetary Systems for the rise of life with SKA). The authors would like to thank Blues Skies Space Limited and UCL for their support of ORBYTS. ORBYTS is a programme to bring researchers to schools to lead original research projects. The work of the students of Hammersmith Academy during ORBYTS formed the basis of this manuscript. The authors thank the anonymous referee for their helpful comments which improved this manuscript.
\software{GILDAS/CLASS http://www.iram.fr/IRAMFR/GILDAS, Emcee \citep{Foreman-Mackey2013}, RADEX \citep{VanderTak2007}}
\appendix
\section{Observed Transitions}
\label{app:tables}
In this section, the observed transitions of CH$_3$OH and CH$_3$CHO are tabulated. Every CH$_3$OH transition with an $E_u < \SI{250}{\kelvin}$ and $A_{ij} > \SI{e-6}{\per\second}$ that fell into the observed frequency range was extracted for processing. For CH$_3$CHO transitions with an E$_u < \SI{100}{\kelvin}$ and $A_{ij} > \SI{e-5}{\per\second}$ were extracted. These ranges were selected to include every 3$\sigma$ detection in the data set and then limit the number of non-detections that contribute to the fit. These are all listed regardless of whether or not they were detected. For non-detections, only the spectroscopic properties of the transition and the integrated emission are presented.\par
Note, the CH$_3$OH transitions between 96.739 and \SI{96.745}{\giga\hertz} and those between 157.271 and \SI{157.276}{\giga\hertz} are blended. The total extent in velocity space of those blended lines are given in the $V_{min}$ and $V_{max}$ columns and thus, the ranges are much broader than the other transitions.
\include{transition_tables}
\section{Spectra}
\label{app:spectra}
\begin{figure*}
\centering
\includegraphics[width=\textwidth]{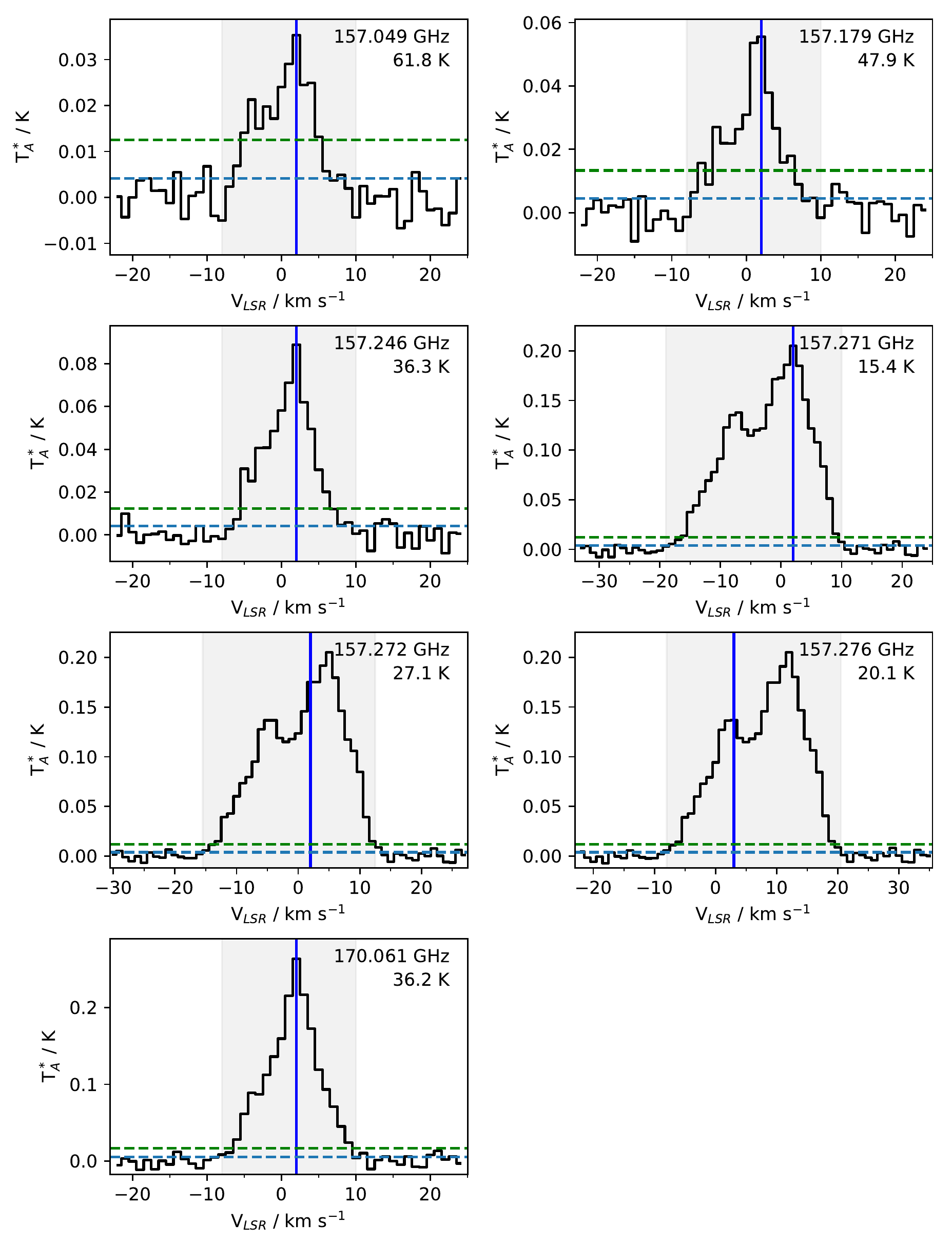}
\caption{Detected transitions of CH$_3$OH in IRAS2A-B. The grey shaded range is the velocity interval over which the emission was integrated. The dashed horizontal lines show the rms noise level and three times that value. The vertical line shows the position of the peak.}
\label{fig:stamps}
\end{figure*}

In this appendix, plots of every detected transition of CH$_3$OH and CH$_3$CHO in each source are presented. The full set is available in the online journal  and an example is shown in Figure~\ref{fig:stamps}. Blue lines indicate velocity at which the transition peaks and grey shaded areas show the interval over which the line was integrated to measure the integrated emission. There are two groups of blended CH$_3$OH lines which can be clearly seen in the spectra. One includes the transitions at 156.271, 156.272 and \SI{157.276}{\giga\hertz} and the other includes the lines at 96.741, 96.745, and \SI{96.756}{\giga\hertz}. For blended lines, the emission was integrated over all blended transitions and the sum of the predicted fluxes was compared to that.
\bibliography{references}



\end{document}

%% file: transition_tables.tex
\begin{table}
\centering
\begin{tabular}{c|ccc|cccccccc}
\toprule
Frequency & E$_u$ & log(A$_{ij}$) & g$_u$ & $\theta_s$ & V$_{min}$& V$_{max}$& V$_{peak}$ & T$_{peak}$ & $\int{T_{MB} dv}$ \\ 
GHz & K & & & " & \si{\kilo\metre\per\second} & \si{\kilo\metre\per\second} & \si{\kilo\metre\per\second} & K & \si{\kelvin\kilo\metre\per\second} \\ 
\midrule
96.739 & 12.54 & -5.59 & 5 & 25.4 & -11.00 & 24.0 & 8.0 & 1.697 (0.006) &  27.08 (5.42)\\ 
96.741 & 6.97 & -5.47 & 5 & 25.4 & -5.00 & 30.0 & 8.0 & 1.651 (0.006) &  26.88 (5.38)\\ 
96.745 & 20.09 & -5.47 & 5 & 25.4 & 5.00 & 40.0 & 9.0 & 0.436 (0.006) &  27.04 (5.41)\\ 
96.756 & 28.01 & -5.58 & 5 & 25.4 & 6.00 & 15.0 & 9.0 & 0.134 (0.006) &  0.89 (0.18)\\ 
97.583 & 21.56 & -5.58 & 5 & 25.2 & 6.00 & 16.0 & 9.0 & 0.219 (0.006) &  1.56 (0.31)\\ 
155.321 & 140.60 & -4.81 & 21 & 15.8 & 8.00 & 10.0 & 9.0 & 0.026 (0.005) &  0.13 (0.03)\\
\bottomrule
\end{tabular}
\caption{Observed transitions of CH$_3$OH in IRAS2A-R. The value of $T_{peak}$ and the velocity measurements of the line are not given where there is no 3$\sigma$ detection. }
\label{table:transitions}
\end{table}
\begin{table}
\centering
\begin{tabular}{c|ccc|cccccccc}
\toprule
Frequency & E$_u$ & log(A$_{ij}$) & g$_u$ & $\theta_s$ & V$_{min}$& V$_{max}$& V$_{peak}$ & T$_{peak}$ & $\int{T_{MB} dv}$ \\ 
GHz & K & & & " & \si{\kilo\metre\per\second} & \si{\kilo\metre\per\second} & \si{\kilo\metre\per\second} & K & \si{\kelvin\kilo\metre\per\second} \\ 
\midrule
155.321 & 140.60 & -4.81 & 21 & 15.8 & - & - & - & - (0.006) &  0.01 (0.02)\\ 
155.998 & 117.46 & -4.78 & 19 & 15.8 & - & - & - & - (0.005) &  0.06 (0.02)\\ 
156.128 & 86.47 & -5.17 & 13 & 15.8 & - & - & - & - (0.005) &  -0.00 (0.02)\\ 
156.489 & 96.61 & -4.75 & 17 & 15.7 & - & - & - & - (0.005) &  0.06 (0.03)\\ 
156.829 & 78.08 & -4.73 & 15 & 15.7 & - & - & - & - (0.006) &  0.13 (0.04)\\ 
157.049 & 61.85 & -4.71 & 13 & 15.7 & -6.00 & 6.0 & 2.0 & 0.035 (0.004) &  0.34 (0.07)\\ 
157.179 & 47.94 & -4.69 & 11 & 15.7 & -5.00 & 7.0 & 2.0 & 0.056 (0.004) &  0.50 (0.10)\\ 
157.246 & 36.34 & -4.68 & 9 & 15.6 & -6.00 & 7.0 & 2.0 & 0.089 (0.004) &  0.77 (0.16)\\ 
157.271 & 15.45 & -4.66 & 3 & 15.6 & -17.00 & 10.0 & 2.0 & 0.205 (0.004) &  3.72 (0.74)\\ 
157.272 & 27.05 & -4.67 & 7 & 15.6 & -14.00 & 13.0 & 2.0 & 0.175 (0.004) &  3.72 (0.74)\\ 
157.276 & 20.09 & -4.66 & 5 & 15.6 & -7.00 & 20.0 & 3.0 & 0.137 (0.004) &  3.72 (0.74)\\ 
167.931 & 125.52 & -4.64 & 19 & 14.6 & - & - & - & - (0.006) &  0.03 (0.03)\\ 
168.578 & 44.26 & -5.36 & 9 & 14.6 & - & - & - & - (0.007) &  0.06 (0.03)\\ 
169.335 & 148.73 & -4.64 & 21 & 14.5 & - & - & - & - (0.006) &  0.02 (0.03)\\ 
170.061 & 36.17 & -4.59 & 7 & 14.5 & -7.00 & 10.0 & 2.0 & 0.263 (0.006) &  2.55 (0.51)\\ 
172.446 & 78.08 & -4.94 & 15 & 14.3 & - & - & - & - (0.010) &  0.11 (0.05)\\ 
\bottomrule
\end{tabular}
\caption{Observed transitions of CH$_3$OH in IRAS2A-B. The value of $T_{peak}$ and the velocity measurements of the line are not given where there is no 3$\sigma$ detection. }
\label{table:transitions}
\end{table}
\begin{table}
\centering
\begin{tabular}{c|ccc|cccccccc}
\toprule
Frequency & E$_u$ & log(A$_{ij}$) & g$_u$ & $\theta_s$ & V$_{min}$& V$_{max}$& V$_{peak}$ & T$_{peak}$ & $\int{T_{MB} dv}$ \\ 
GHz & K & & & " & \si{\kilo\metre\per\second} & \si{\kilo\metre\per\second} & \si{\kilo\metre\per\second} & K & \si{\kelvin\kilo\metre\per\second} \\ 
\midrule
96.739 & 12.54 & -5.59 & 5 & 25.4 & -11.00 & 17.0 & 8.0 & 1.066 (0.005) &  6.84 (1.37)\\ 
96.741 & 6.97 & -5.47 & 5 & 25.4 & -5.00 & 9.0 & 8.0 & 1.337 (0.006) &  6.75 (1.35)\\ 
96.745 & 20.09 & -5.47 & 5 & 25.4 & 5.00 & 19.0 & 8.0 & 0.252 (0.005) &  6.84 (1.37)\\ 
96.756 & 28.01 & -5.58 & 5 & 25.4 & 6.00 & 9.0 & 8.0 & 0.063 (0.005) &  0.18 (0.04)\\ 
97.583 & 21.56 & -5.58 & 5 & 25.2 & 6.00 & 9.0 & 7.0 & 0.061 (0.006) &  0.21 (0.05)\\ 
155.321 & 140.60 & -4.81 & 21 & 15.8 & - & - & - & - (0.005) &  0.04 (0.02)\\
\bottomrule
\end{tabular}
\caption{Observed transitions of CH$_3$OH in IRAS4A-R. The value of $T_{peak}$ and the velocity measurements of the line are not given where there is no 3$\sigma$ detection. }
\label{table:transitions}
\end{table}
\begin{table}
\centering
\begin{tabular}{c|ccc|cccccccc}
\toprule
Frequency & E$_u$ & log(A$_{ij}$) & g$_u$ & $\theta_s$ & V$_{min}$& V$_{max}$& V$_{peak}$ & T$_{peak}$ & $\int{T_{MB} dv}$ \\ 
GHz & K & & & " & \si{\kilo\metre\per\second} & \si{\kilo\metre\per\second} & \si{\kilo\metre\per\second} & K & \si{\kelvin\kilo\metre\per\second} \\ 
\midrule
96.739 & 12.54 & -5.59 & 5 & 25.4 & -19.00 & 9.0 & 6.0 & 0.824 (0.019) &  19.74 (3.95)\\ 
96.741 & 6.97 & -5.47 & 5 & 25.4 & -13.00 & 15.0 & 6.0 & 1.336 (0.021) &  19.60 (3.92)\\ 
96.745 & 20.09 & -5.47 & 5 & 25.4 & -3.00 & 25.0 & 6.0 & 0.501 (0.022) &  19.98 (4.00)\\ 
96.756 & 28.01 & -5.58 & 5 & 25.4 & -4.00 & 9.0 & 6.0 & 0.113 (0.005) &  0.83 (0.17)\\ 
97.583 & 21.56 & -5.58 & 5 & 25.2 & -4.00 & 9.0 & 6.0 & 0.177 (0.005) &  1.35 (0.27)\\ 
99.611 & 61.85 & -9.83 & 13 & 24.7 & - & - & - & - (0.005) &  0.04 (0.03)\\
\bottomrule
\end{tabular}
\caption{Observed transitions of CH$_3$OH in IRAS4A-B. The value of $T_{peak}$ and the velocity measurements of the line are not given where there is no 3$\sigma$ detection. }
\label{table:transitions}
\end{table}
\begin{table}
\centering
\begin{tabular}{c|ccc|cccccccc}
\toprule
Frequency & E$_u$ & log(A$_{ij}$) & g$_u$ & $\theta_s$ & V$_{min}$& V$_{max}$& V$_{peak}$ & T$_{peak}$ & $\int{T_{MB} dv}$ \\ 
GHz & K & & & " & \si{\kilo\metre\per\second} & \si{\kilo\metre\per\second} & \si{\kilo\metre\per\second} & K & \si{\kelvin\kilo\metre\per\second} \\ 
\midrule
167.931 & 125.52 & -4.64 & 19 & 14.6 & 3.00 & 5.0 & 4.0 & 0.019 (0.005) &  0.11 (0.03)\\ 
168.464 & 112.71 & -7.91 & 15 & 14.6 & - & - & - & - (0.005) &  -0.01 (0.03)\\ 
168.578 & 44.26 & -5.36 & 9 & 14.6 & 3.00 & 6.0 & 4.0 & 0.028 (0.006) &  0.28 (0.07)\\ 
169.335 & 148.73 & -4.64 & 21 & 14.5 & - & - & - & - (0.007) &  0.05 (0.04)\\ 
170.061 & 36.17 & -4.59 & 7 & 14.5 & -5.00 & 9.0 & 6.0 & 0.341 (0.007) &  3.63 (0.73)\\ 
172.446 & 78.08 & -4.94 & 15 & 14.3 & 4.00 & 7.0 & 6.0 & 0.036 (0.009) &  0.45 (0.10)\\ 
\bottomrule
\end{tabular}
\caption{Observed transitions of CH$_3$OH in IRAS4A-B. The value of $T_{peak}$ and the velocity measurements of the line are not given where there is no 3$\sigma$ detection. continued.}
\label{table:transitions}
\end{table}
\begin{table}
\centering
\begin{tabular}{c|ccc|cccccccc}
\toprule
Frequency & E$_u$ & log(A$_{ij}$) & g$_u$ & $\theta_s$ & V$_{min}$& V$_{max}$& V$_{peak}$ & T$_{peak}$ & $\int{T_{MB} dv}$ \\ 
GHz & K & & & " & \si{\kilo\metre\per\second} & \si{\kilo\metre\per\second} & \si{\kilo\metre\per\second} & K & \si{\kelvin\kilo\metre\per\second} \\ 
\midrule
96.739 & 12.54 & -5.59 & 5 & 25.4 & -4.00 & 18.0 & 17.0 & 0.030 (0.007) &  1.55 (0.31)\\ 
96.741 & 6.97 & -5.47 & 5 & 25.4 & 2.00 & 25.0 & 17.0 & 0.065 (0.005) &  1.49 (0.30)\\ 
96.745 & 20.09 & -5.47 & 5 & 25.4 & 12.00 & 34.0 & 17.0 & 0.054 (0.006) &  1.45 (0.29)\\ 
96.756 & 28.01 & -5.58 & 5 & 25.4 & - & - & - & - (0.006) &  -0.03 (0.02)\\ 
97.583 & 21.56 & -5.58 & 5 & 25.2 & - & - & - & - (0.008) &  0.00 (0.02)\\ 
155.321 & 140.60 & -4.81 & 21 & 15.8 & - & - & - & - (0.004) &  0.02 (0.01)\\
\bottomrule
\end{tabular}
\caption{Observed transitions of CH$_3$OH in L1157-R. The value of $T_{peak}$ and the velocity measurements of the line are not given where there is no 3$\sigma$ detection. }
\label{table:transitions}
\end{table}
\begin{table}
\centering
\begin{tabular}{c|ccc|cccccccc}
\toprule
Frequency & E$_u$ & log(A$_{ij}$) & g$_u$ & $\theta_s$ & V$_{min}$& V$_{max}$& V$_{peak}$ & T$_{peak}$ & $\int{T_{MB} dv}$ \\ 
GHz & K & & & " & \si{\kilo\metre\per\second} & \si{\kilo\metre\per\second} & \si{\kilo\metre\per\second} & K & \si{\kelvin\kilo\metre\per\second} \\ 
\midrule
96.739 & 12.54 & -5.59 & 5 & 25.4 & -20.00 & 7.0 & 2.0 & 2.659 (0.004) &  30.36 (6.07)\\ 
96.741 & 6.97 & -5.47 & 5 & 25.4 & -14.00 & 14.0 & 2.0 & 3.378 (0.004) &  30.38 (6.08)\\ 
96.745 & 20.09 & -5.47 & 5 & 25.4 & -4.00 & 23.0 & 2.0 & 0.699 (0.005) &  30.40 (6.08)\\ 
96.756 & 28.01 & -5.58 & 5 & 25.4 & -3.00 & 4.0 & 2.0 & 0.192 (0.005) &  0.65 (0.13)\\ 
97.583 & 21.56 & -5.58 & 5 & 25.2 & -3.00 & 5.0 & 2.0 & 0.307 (0.004) &  1.12 (0.22)\\ 
155.321 & 140.60 & -4.81 & 21 & 15.8 & - & - & - & - (0.004) &  0.01 (0.01)\\
\bottomrule
\end{tabular}
\caption{Observed transitions of CH$_3$OH in L1157-B2. The value of $T_{peak}$ and the velocity measurements of the line are not given where there is no 3$\sigma$ detection. }
\label{table:transitions}
\end{table}
\begin{table}
\centering
\begin{tabular}{c|ccc|cccccccc}
\toprule
Frequency & E$_u$ & log(A$_{ij}$) & g$_u$ & $\theta_s$ & V$_{min}$& V$_{max}$& V$_{peak}$ & T$_{peak}$ & $\int{T_{MB} dv}$ \\ 
GHz & K & & & " & \si{\kilo\metre\per\second} & \si{\kilo\metre\per\second} & \si{\kilo\metre\per\second} & K & \si{\kelvin\kilo\metre\per\second} \\ 
\midrule
155.321 & 140.60 & -4.81 & 21 & 15.8 & - & - & - & - (0.007) &  0.02 (0.01)\\ 
155.998 & 117.46 & -4.78 & 19 & 15.8 & - & - & - & - (0.006) &  -0.01 (0.01)\\ 
156.128 & 86.47 & -5.17 & 13 & 15.8 & - & - & - & - (0.007) &  0.01 (0.01)\\ 
156.489 & 96.61 & -4.75 & 17 & 15.7 & - & - & - & - (0.004) &  -0.01 (0.01)\\ 
156.602 & 21.44 & -5.05 & 10 & 15.7 & - & - & - & - (0.004) &  0.00 (0.01)\\ 
156.829 & 78.08 & -4.73 & 15 & 15.7 & - & - & - & - (0.005) &  -0.01 (0.01)\\ 
157.049 & 61.85 & -4.71 & 13 & 15.7 & - & - & - & - (0.004) &  -0.00 (0.01)\\ 
157.179 & 47.94 & -4.69 & 11 & 15.7 & - & - & - & - (0.004) &  -0.00 (0.01)\\ 
157.246 & 36.34 & -4.68 & 9 & 15.6 & - & - & - & - (0.005) &  0.01 (0.01)\\ 
157.271 & 15.45 & -4.66 & 3 & 15.6 & - & - & - & - (0.004) &  0.17 (0.04)\\ 
157.272 & 27.05 & -4.67 & 7 & 15.6 & 6.00 & 9.0 & 7.0 & 0.052 (0.004) &  0.18 (0.04)\\ 
157.276 & 20.09 & -4.66 & 5 & 15.6 & - & - & - & - (0.004) &  0.16 (0.03)\\ 
168.578 & 44.26 & -5.36 & 9 & 14.6 & - & - & - & - (0.006) &  -0.01 (0.01)\\ 
169.335 & 148.73 & -4.64 & 21 & 14.5 & - & - & - & - (0.006) &  0.01 (0.01)\\ 
170.061 & 36.17 & -4.59 & 7 & 14.5 & - & - & - & - (0.007) &  -0.02 (0.01)\\ 
172.446 & 78.08 & -4.94 & 15 & 14.3 & - & - & - & - (0.007) &  0.00 (0.01)\\ 
\bottomrule
\end{tabular}
\caption{Observed transitions of CH$_3$OH in L1448-B2. The value of $T_{peak}$ and the velocity measurements of the line are not given where there is no 3$\sigma$ detection. }
\label{table:transitions}
\end{table}
\begin{table}
\centering
\begin{tabular}{c|ccc|cccccccc}
\toprule
Frequency & E$_u$ & log(A$_{ij}$) & g$_u$ & $\theta_s$ & V$_{min}$& V$_{max}$& V$_{peak}$ & T$_{peak}$ & $\int{T_{MB} dv}$ \\ 
GHz & K & & & " & \si{\kilo\metre\per\second} & \si{\kilo\metre\per\second} & \si{\kilo\metre\per\second} & K & \si{\kelvin\kilo\metre\per\second} \\ 
\midrule
96.274 & 22.93 & -4.60 & 22 & 25.6 & - & - & - & - (0.007) &  0.06 (0.02)\\ 
96.343 & 50.06 & -4.97 & 22 & 25.5 & - & - & - & - (0.006) &  -0.02 (0.02)\\ 
96.343 & 50.06 & -4.97 & 22 & 25.5 & - & - & - & - (0.006) &  -0.02 (0.02)\\ 
96.353 & 49.99 & -4.97 & 22 & 25.5 & - & - & - & - (0.006) &  -0.02 (0.02)\\ 
96.361 & 49.95 & -4.97 & 22 & 25.5 & - & - & - & - (0.006) &  -0.00 (0.02)\\ 
96.368 & 34.26 & -4.72 & 22 & 25.5 & - & - & - & - (0.007) &  0.03 (0.02)\\
\bottomrule
\end{tabular}
\caption{Observed transitions of CH$_3$CHO in IRAS2A-R. The value of $T_{peak}$ and the velocity measurements of the line are not given where there is no 3$\sigma$ detection. }
\label{table:transitions}
\end{table}
\begin{table}
\centering
\begin{tabular}{c|ccc|cccccccc}
\toprule
Frequency & E$_u$ & log(A$_{ij}$) & g$_u$ & $\theta_s$ & V$_{min}$& V$_{max}$& V$_{peak}$ & T$_{peak}$ & $\int{T_{MB} dv}$ \\ 
GHz & K & & & " & \si{\kilo\metre\per\second} & \si{\kilo\metre\per\second} & \si{\kilo\metre\per\second} & K & \si{\kelvin\kilo\metre\per\second} \\ 
\midrule
162.044 & 50.44 & -4.79 & 42 & 15.2 & - & - & - & - (0.008) &  0.00 (0.02)\\ 
164.727 & 50.64 & -4.90 & 38 & 14.9 & - & - & - & - (0.008) &  -0.02 (0.02)\\ 
167.981 & 27.42 & -4.74 & 30 & 14.6 & - & - & - & - (0.006) &  0.04 (0.02)\\ 
168.089 & 42.73 & -3.79 & 38 & 14.6 & 6.00 & 13.0 & 9.0 & 0.031 (0.006) &  0.20 (0.04)\\ 
168.093 & 42.66 & -3.79 & 38 & 14.6 & 6.00 & 10.0 & 7.0 & 0.031 (0.006) &  0.11 (0.03)\\ 
170.293 & 59.86 & -4.86 & 42 & 14.4 & - & - & - & - (0.008) &  -0.03 (0.02)\\
\bottomrule
\end{tabular}
\caption{Observed transitions of CH$_3$CHO in IRAS2A-R. The value of $T_{peak}$ and the velocity measurements of the line are not given where there is no 3$\sigma$ detection. continued.}
\label{table:transitions}
\end{table}
\begin{table}
\centering
\begin{tabular}{c|ccc|cccccccc}
\toprule
Frequency & E$_u$ & log(A$_{ij}$) & g$_u$ & $\theta_s$ & V$_{min}$& V$_{max}$& V$_{peak}$ & T$_{peak}$ & $\int{T_{MB} dv}$ \\ 
GHz & K & & & " & \si{\kilo\metre\per\second} & \si{\kilo\metre\per\second} & \si{\kilo\metre\per\second} & K & \si{\kelvin\kilo\metre\per\second} \\ 
\midrule
152.635 & 33.10 & -3.91 & 34 & 16.1 & - & - & - & - (0.005) &  0.03 (0.02)\\ 
153.004 & 28.48 & -4.97 & 26 & 16.1 & - & - & - & - (0.006) &  -0.01 (0.02)\\ 
153.873 & 42.33 & -3.93 & 34 & 16.0 & - & - & - & - (0.006) &  -0.05 (0.03)\\ 
154.027 & 42.33 & -3.93 & 34 & 16.0 & - & - & - & - (0.006) &  0.06 (0.03)\\ 
154.161 & 89.76 & -4.11 & 34 & 16.0 & - & - & - & - (0.005) &  -0.01 (0.02)\\ 
154.161 & 89.76 & -4.11 & 34 & 16.0 & - & - & - & - (0.005) &  -0.01 (0.02)\\
\bottomrule
\end{tabular}
\caption{Observed transitions of CH$_3$CHO in IRAS2A-B. The value of $T_{peak}$ and the velocity measurements of the line are not given where there is no 3$\sigma$ detection. }
\label{table:transitions}
\end{table}
\begin{table}
\centering
\begin{tabular}{c|ccc|cccccccc}
\toprule
Frequency & E$_u$ & log(A$_{ij}$) & g$_u$ & $\theta_s$ & V$_{min}$& V$_{max}$& V$_{peak}$ & T$_{peak}$ & $\int{T_{MB} dv}$ \\ 
GHz & K & & & " & \si{\kilo\metre\per\second} & \si{\kilo\metre\per\second} & \si{\kilo\metre\per\second} & K & \si{\kelvin\kilo\metre\per\second} \\ 
\midrule
173.638 & 62.03 & -3.79 & 38 & 14.2 & - & - & - & - (0.009) &  0.01 (0.04)\\ 
173.664 & 61.93 & -3.79 & 38 & 14.2 & - & - & - & - (0.009) &  0.07 (0.04)\\ 
174.982 & 50.94 & -3.75 & 38 & 14.1 & - & - & - & - (0.012) &  0.04 (0.05)\\ 
175.091 & 50.91 & -3.75 & 38 & 14.0 & - & - & - & - (0.011) &  0.06 (0.05)\\ 
\bottomrule
\end{tabular}
\caption{Observed transitions of CH$_3$CHO in IRAS2A-B. The value of $T_{peak}$ and the velocity measurements of the line are not given where there is no 3$\sigma$ detection. continued.}
\label{table:transitions}
\end{table}
\begin{table}
\centering
\begin{tabular}{c|ccc|cccccccc}
\toprule
Frequency & E$_u$ & log(A$_{ij}$) & g$_u$ & $\theta_s$ & V$_{min}$& V$_{max}$& V$_{peak}$ & T$_{peak}$ & $\int{T_{MB} dv}$ \\ 
GHz & K & & & " & \si{\kilo\metre\per\second} & \si{\kilo\metre\per\second} & \si{\kilo\metre\per\second} & K & \si{\kelvin\kilo\metre\per\second} \\ 
\midrule
96.274 & 22.93 & -4.60 & 22 & 25.6 & - & - & - & - (0.006) &  0.02 (0.01)\\ 
96.343 & 50.06 & -4.97 & 22 & 25.5 & - & - & - & - (0.006) &  0.00 (0.01)\\ 
96.343 & 50.06 & -4.97 & 22 & 25.5 & - & - & - & - (0.006) &  0.00 (0.01)\\ 
96.353 & 49.99 & -4.97 & 22 & 25.5 & - & - & - & - (0.007) &  -0.00 (0.01)\\ 
96.361 & 49.95 & -4.97 & 22 & 25.5 & - & - & - & - (0.007) &  0.01 (0.01)\\ 
96.368 & 34.26 & -4.72 & 22 & 25.5 & - & - & - & - (0.007) &  0.02 (0.01)\\
\bottomrule
\end{tabular}
\caption{Observed transitions of CH$_3$CHO in IRAS4A-R. The value of $T_{peak}$ and the velocity measurements of the line are not given where there is no 3$\sigma$ detection. }
\label{table:transitions}
\end{table}
\begin{table}
\centering
\begin{tabular}{c|ccc|cccccccc}
\toprule
Frequency & E$_u$ & log(A$_{ij}$) & g$_u$ & $\theta_s$ & V$_{min}$& V$_{max}$& V$_{peak}$ & T$_{peak}$ & $\int{T_{MB} dv}$ \\ 
GHz & K & & & " & \si{\kilo\metre\per\second} & \si{\kilo\metre\per\second} & \si{\kilo\metre\per\second} & K & \si{\kelvin\kilo\metre\per\second} \\ 
\midrule
162.044 & 50.44 & -4.79 & 42 & 15.2 & - & - & - & - (0.008) &  -0.00 (0.01)\\ 
164.727 & 50.64 & -4.90 & 38 & 14.9 & - & - & - & - (0.011) &  -0.01 (0.02)\\ 
167.981 & 27.42 & -4.74 & 30 & 14.6 & - & - & - & - (0.011) &  -0.02 (0.02)\\ 
168.089 & 42.73 & -3.79 & 38 & 14.6 & - & - & - & - (0.012) &  0.00 (0.02)\\ 
168.093 & 42.66 & -3.79 & 38 & 14.6 & - & - & - & - (0.010) &  0.04 (0.02)\\ 
170.293 & 59.86 & -4.86 & 42 & 14.4 & - & - & - & - (0.007) &  -0.01 (0.01)\\
\bottomrule
\end{tabular}
\caption{Observed transitions of CH$_3$CHO in IRAS4A-R. The value of $T_{peak}$ and the velocity measurements of the line are not given where there is no 3$\sigma$ detection. continued.}
\label{table:transitions}
\end{table}
\begin{table}
\centering
\begin{tabular}{c|ccc|cccccccc}
\toprule
Frequency & E$_u$ & log(A$_{ij}$) & g$_u$ & $\theta_s$ & V$_{min}$& V$_{max}$& V$_{peak}$ & T$_{peak}$ & $\int{T_{MB} dv}$ \\ 
GHz & K & & & " & \si{\kilo\metre\per\second} & \si{\kilo\metre\per\second} & \si{\kilo\metre\per\second} & K & \si{\kelvin\kilo\metre\per\second} \\ 
\midrule
96.274 & 22.93 & -4.60 & 22 & 25.6 & - & - & - & - (0.006) &  0.10 (0.03)\\ 
96.343 & 50.06 & -4.97 & 22 & 25.5 & - & - & - & - (0.006) &  0.02 (0.02)\\ 
96.343 & 50.06 & -4.97 & 22 & 25.5 & - & - & - & - (0.006) &  0.02 (0.02)\\ 
96.353 & 49.99 & -4.97 & 22 & 25.5 & - & - & - & - (0.006) &  0.03 (0.02)\\ 
96.361 & 49.95 & -4.97 & 22 & 25.5 & - & - & - & - (0.005) &  -0.02 (0.02)\\ 
96.368 & 34.26 & -4.72 & 22 & 25.5 & - & - & - & - (0.005) &  0.08 (0.02)\\
\bottomrule
\end{tabular}
\caption{Observed transitions of CH$_3$CHO in IRAS4A-B. The value of $T_{peak}$ and the velocity measurements of the line are not given where there is no 3$\sigma$ detection. }
\label{table:transitions}
\end{table}
\begin{table}
\centering
\begin{tabular}{c|ccc|cccccccc}
\toprule
Frequency & E$_u$ & log(A$_{ij}$) & g$_u$ & $\theta_s$ & V$_{min}$& V$_{max}$& V$_{peak}$ & T$_{peak}$ & $\int{T_{MB} dv}$ \\ 
GHz & K & & & " & \si{\kilo\metre\per\second} & \si{\kilo\metre\per\second} & \si{\kilo\metre\per\second} & K & \si{\kelvin\kilo\metre\per\second} \\ 
\midrule
164.727 & 50.64 & -4.90 & 38 & 14.9 & - & - & - & - (0.009) &  -0.03 (0.03)\\ 
167.981 & 27.42 & -4.74 & 30 & 14.6 & - & - & - & - (0.004) &  0.01 (0.02)\\ 
168.089 & 42.73 & -3.79 & 38 & 14.6 & 6.00 & 8.0 & 7.0 & 0.021 (0.005) &  0.14 (0.03)\\ 
168.093 & 42.66 & -3.79 & 38 & 14.6 & 5.00 & 8.0 & 6.0 & 0.021 (0.005) &  0.08 (0.02)\\ 
170.293 & 59.86 & -4.86 & 42 & 14.4 & - & - & - & - (0.008) &  -0.03 (0.03)\\ 
171.265 & 41.42 & -3.76 & 38 & 14.4 & - & - & - & - (0.009) &  0.11 (0.04)\\
\bottomrule
\end{tabular}
\caption{Observed transitions of CH$_3$CHO in IRAS4A-B. The value of $T_{peak}$ and the velocity measurements of the line are not given where there is no 3$\sigma$ detection. continued.}
\label{table:transitions}
\end{table}
\begin{table}
\centering
\begin{tabular}{c|ccc|cccccccc}
\toprule
Frequency & E$_u$ & log(A$_{ij}$) & g$_u$ & $\theta_s$ & V$_{min}$& V$_{max}$& V$_{peak}$ & T$_{peak}$ & $\int{T_{MB} dv}$ \\ 
GHz & K & & & " & \si{\kilo\metre\per\second} & \si{\kilo\metre\per\second} & \si{\kilo\metre\per\second} & K & \si{\kelvin\kilo\metre\per\second} \\ 
\midrule
96.274 & 22.93 & -4.60 & 22 & 25.6 & - & - & - & - (0.007) &  -0.03 (0.02)\\ 
96.343 & 50.06 & -4.97 & 22 & 25.5 & - & - & - & - (0.006) &  0.02 (0.02)\\ 
96.343 & 50.06 & -4.97 & 22 & 25.5 & - & - & - & - (0.006) &  0.02 (0.02)\\ 
96.353 & 49.99 & -4.97 & 22 & 25.5 & - & - & - & - (0.006) &  0.02 (0.02)\\ 
96.361 & 49.95 & -4.97 & 22 & 25.5 & - & - & - & - (0.007) &  0.00 (0.02)\\ 
96.368 & 34.26 & -4.72 & 22 & 25.5 & - & - & - & - (0.007) &  -0.01 (0.02)\\
\bottomrule
\end{tabular}
\caption{Observed transitions of CH$_3$CHO in L1157-R. The value of $T_{peak}$ and the velocity measurements of the line are not given where there is no 3$\sigma$ detection. }
\label{table:transitions}
\end{table}
\begin{table}
\centering
\begin{tabular}{c|ccc|cccccccc}
\toprule
Frequency & E$_u$ & log(A$_{ij}$) & g$_u$ & $\theta_s$ & V$_{min}$& V$_{max}$& V$_{peak}$ & T$_{peak}$ & $\int{T_{MB} dv}$ \\ 
GHz & K & & & " & \si{\kilo\metre\per\second} & \si{\kilo\metre\per\second} & \si{\kilo\metre\per\second} & K & \si{\kelvin\kilo\metre\per\second} \\ 
\midrule
162.044 & 50.44 & -4.79 & 42 & 15.2 & - & - & - & - (0.007) &  -0.01 (0.02)\\ 
164.727 & 50.64 & -4.90 & 38 & 14.9 & - & - & - & - (0.011) &  -0.00 (0.03)\\ 
167.981 & 27.42 & -4.74 & 30 & 14.6 & - & - & - & - (0.005) &  -0.00 (0.02)\\ 
168.089 & 42.73 & -3.79 & 38 & 14.6 & - & - & - & - (0.005) &  0.01 (0.01)\\ 
168.093 & 42.66 & -3.79 & 38 & 14.6 & - & - & - & - (0.005) &  0.01 (0.02)\\ 
170.293 & 59.86 & -4.86 & 42 & 14.4 & - & - & - & - (0.005) &  -0.02 (0.02)\\
\bottomrule
\end{tabular}
\caption{Observed transitions of CH$_3$CHO in L1157-R. The value of $T_{peak}$ and the velocity measurements of the line are not given where there is no 3$\sigma$ detection. continued.}
\label{table:transitions}
\end{table}
\begin{table}
\centering
\begin{tabular}{c|ccc|cccccccc}
\toprule
Frequency & E$_u$ & log(A$_{ij}$) & g$_u$ & $\theta_s$ & V$_{min}$& V$_{max}$& V$_{peak}$ & T$_{peak}$ & $\int{T_{MB} dv}$ \\ 
GHz & K & & & " & \si{\kilo\metre\per\second} & \si{\kilo\metre\per\second} & \si{\kilo\metre\per\second} & K & \si{\kelvin\kilo\metre\per\second} \\ 
\midrule
96.274 & 22.93 & -4.60 & 22 & 25.6 & 1.00 & 3.0 & 2.0 & 0.020 (0.005) &  0.04 (0.01)\\ 
96.343 & 50.06 & -4.97 & 22 & 25.5 & - & - & - & - (0.005) &  0.00 (0.01)\\ 
96.343 & 50.06 & -4.97 & 22 & 25.5 & - & - & - & - (0.005) &  0.00 (0.01)\\ 
96.353 & 49.99 & -4.97 & 22 & 25.5 & - & - & - & - (0.005) &  0.00 (0.01)\\ 
96.361 & 49.95 & -4.97 & 22 & 25.5 & - & - & - & - (0.005) &  0.02 (0.01)\\ 
96.368 & 34.26 & -4.72 & 22 & 25.5 & - & - & - & - (0.004) &  0.01 (0.01)\\
\bottomrule
\end{tabular}
\caption{Observed transitions of CH$_3$CHO in L1157-B2. The value of $T_{peak}$ and the velocity measurements of the line are not given where there is no 3$\sigma$ detection. }
\label{table:transitions}
\end{table}
\begin{table}
\centering
\begin{tabular}{c|ccc|cccccccc}
\toprule
Frequency & E$_u$ & log(A$_{ij}$) & g$_u$ & $\theta_s$ & V$_{min}$& V$_{max}$& V$_{peak}$ & T$_{peak}$ & $\int{T_{MB} dv}$ \\ 
GHz & K & & & " & \si{\kilo\metre\per\second} & \si{\kilo\metre\per\second} & \si{\kilo\metre\per\second} & K & \si{\kelvin\kilo\metre\per\second} \\ 
\midrule
162.044 & 50.44 & -4.79 & 42 & 15.2 & - & - & - & - (0.005) &  -0.00 (0.01)\\ 
164.727 & 50.64 & -4.90 & 38 & 14.9 & - & - & - & - (0.008) &  -0.00 (0.02)\\ 
167.981 & 27.42 & -4.74 & 30 & 14.6 & - & - & - & - (0.005) &  0.02 (0.01)\\ 
168.089 & 42.73 & -3.79 & 38 & 14.6 & - & - & - & - (0.006) &  0.05 (0.02)\\ 
168.093 & 42.66 & -3.79 & 38 & 14.6 & - & - & - & - (0.005) &  0.03 (0.01)\\ 
170.293 & 59.86 & -4.86 & 42 & 14.4 & - & - & - & - (0.004) &  -0.01 (0.01)\\
\bottomrule
\end{tabular}
\caption{Observed transitions of CH$_3$CHO in L1157-B2. The value of $T_{peak}$ and the velocity measurements of the line are not given where there is no 3$\sigma$ detection. continued.}
\label{table:transitions}
\end{table}
\begin{table}
\centering
\begin{tabular}{c|ccc|cccccccc}
\toprule
Frequency & E$_u$ & log(A$_{ij}$) & g$_u$ & $\theta_s$ & V$_{min}$& V$_{max}$& V$_{peak}$ & T$_{peak}$ & $\int{T_{MB} dv}$ \\ 
GHz & K & & & " & \si{\kilo\metre\per\second} & \si{\kilo\metre\per\second} & \si{\kilo\metre\per\second} & K & \si{\kelvin\kilo\metre\per\second} \\ 
\midrule
152.635 & 33.10 & -3.91 & 34 & 16.1 & - & - & - & - (0.005) &  -0.01 (0.01)\\ 
153.004 & 28.48 & -4.97 & 26 & 16.1 & - & - & - & - (0.006) &  0.01 (0.01)\\ 
153.873 & 42.33 & -3.93 & 34 & 16.0 & - & - & - & - (0.006) &  -0.01 (0.01)\\ 
154.027 & 42.33 & -3.93 & 34 & 16.0 & - & - & - & - (0.006) &  0.01 (0.01)\\ 
154.161 & 89.76 & -4.11 & 34 & 16.0 & - & - & - & - (0.006) &  0.01 (0.01)\\ 
154.161 & 89.76 & -4.11 & 34 & 16.0 & - & - & - & - (0.006) &  0.01 (0.01)\\
\bottomrule
\end{tabular}
\caption{Observed transitions of CH$_3$CHO in L1448-B2. The value of $T_{peak}$ and the velocity measurements of the line are not given where there is no 3$\sigma$ detection. }
\label{table:transitions}
\end{table}
\begin{table}
\centering
\begin{tabular}{c|ccc|cccccccc}
\toprule
Frequency & E$_u$ & log(A$_{ij}$) & g$_u$ & $\theta_s$ & V$_{min}$& V$_{max}$& V$_{peak}$ & T$_{peak}$ & $\int{T_{MB} dv}$ \\ 
GHz & K & & & " & \si{\kilo\metre\per\second} & \si{\kilo\metre\per\second} & \si{\kilo\metre\per\second} & K & \si{\kelvin\kilo\metre\per\second} \\ 
\midrule
174.982 & 50.94 & -3.75 & 38 & 14.1 & - & - & - & - (0.008) &  -0.00 (0.01)\\ 
175.091 & 50.91 & -3.75 & 38 & 14.0 & - & - & - & - (0.010) &  -0.01 (0.02)\\ 
\bottomrule
\end{tabular}
\caption{Observed transitions of CH$_3$CHO in L1448-B2. The value of $T_{peak}$ and the velocity measurements of the line are not given where there is no 3$\sigma$ detection. continued.}
\label{table:transitions}
\end{table}